\documentclass[printer]{aa}
\usepackage{graphicx}
\usepackage{txfonts}
\def\lesssim{\mathrel{\hbox{\rlap{\hbox{\lower4pt\hbox{$\sim$}}}\hbox{$<$}}}}
\def\gtrsim{\mathrel{\hbox{\rlap{\hbox{\lower4pt\hbox{$\sim$}}}\hbox{$>$}}}}
\newcommand{\mincir}{\raise -2.truept\hbox{\rlap{\hbox{$\sim$}}\raise5.truept
\hbox{$<$}\ }}
\newcommand{\magcir}{\raise -2.truept\hbox{\rlap{\hbox{$\sim$}}\raise5.truept
\hbox{$>$}\ }}
\newcommand{\siml}{\raise -2.truept\hbox{\rlap{\hbox{$\sim$}}\raise5.truept
\hbox{$<$}\ }}
\newcommand{\simg}{\raise -2.truept\hbox{\rlap{\hbox{$\sim$}}\raise5.truept
\hbox{$>$}\ }}
\newcommand{\be}{\begin{equation}}
\newcommand{\ee}{\end{equation}}
\newcommand{\ba}{\begin{eqnarray}}
\newcommand{\ea}{\end{eqnarray}}
\newcommand {\kpc} {$\mathrm{h_{70}^{-1}}$ kpc$\;$}
\newcommand {\kpcc} {$\mathrm{h_{70}^{-1}}$ kpc}
\newcommand {\h} {$\mathrm{h_{70}^{-1}}$ Mpc$\;$}
\newcommand {\hh} {$\mathrm{h_{70}^{-1}}$ Mpc}
\newcommand {\hhh} {\;\mathrm{h_{70}^{-1}} \mathrm{Mpc}}
\newcommand {\ks} {km~s$^{-1} \;$}
\newcommand {\kss} {km~s$^{-1}$}

\newcommand {\mqua} {$\times 10^{14}\;\mathrm{h_{70}^{-1}}\;M_{\odot} \;$}
\newcommand {\mquaa} {$\times 10^{14}\;\mathrm{h_{70}^{-1}}\;M_{\odot}$}
\newcommand {\mqui} {$\times 10^{15}\;\mathrm{h_{70}^{-1}}\;M_{\odot} \;$}
\newcommand {\mquii} {$\times 10^{15}\;\mathrm{h_{70}^{-1}}\;M_{\odot}$}

\newcommand {\mll} {$\mathrm{h_{70}}\;M_{\odot}/L_{\odot}$}

\newcommand{\degree}{\ensuremath{\mathrm{^\circ}}}
\newcommand{\arcm}{\ensuremath{\mathrm{^\prime}\;}}
\newcommand{\arcs}{\ensuremath{\arcmm\hskip -0.1em\arcmm \;}}
\newcommand{\arcmm}{\ensuremath{\mathrm{^\prime}}}
\newcommand{\arcss}{\ensuremath{\arcmm\hskip -0.1em\arcmm}}
\newcommand{\dotarcs}{\,\rlap{\hbox{$\mathrm{^\prime\hskip-0.1em^\prime}$}}{\hbox{$.$}}\,}

\newcommand{\dotsec}{\,\rlap{\hbox{$\mathrm{^s}$}}{\hbox{$.$}}\,}
\begin{document}
   \title{Internal dynamics of the massive cluster Abell 697: \\
a  multiwavelength analysis}
%
\author{M. Girardi\inst{1,2}
\and W. Boschin\inst{1,3} 
\and R. Barrena\inst{4}}
   \offprints{M. Girardi; e.mail: girardi@oats.inaf.it}

\institute{
Dipartimento di Astronomia, Universit\`{a} degli Studi di Trieste, via Tiepolo 11, I-34131 Trieste, Italy\\
\and
INAF -- Osservatorio Astronomico di Trieste, via Tiepolo 11, I-34131  Trieste, Italy\\
\and 
Fundaci\'on Galileo Galilei - INAF, C/Alvarez de Abreu 70, E-38700 Santa Cruz de La Palma, Canary Islands, Spain\\
\and
Instituto de Astrofisica de Canarias, C/Via Lactea s/n, E-38200 La Laguna, Tenerife, Canary Islands, Spain\\
}

\date{Received / accepted }

\abstract{}{We conduct an intensive study of the rich, X--ray
luminous, and hot galaxy cluster Abell 697 (at $z=0.282$), likely
containing a diffuse radio--emission, to determine its dynamical
status.}{Our analysis is based on new spectroscopic data obtained at
the TNG telescope for 93 galaxies and on new photometric data obtained
at the INT telescope in a large field.  We combine galaxy velocity and
position information to select 68 cluster members (out to $\sim 1.3$
\h from the cD galaxy), determine global dynamical properties, and
detect possible substructures. The investigation of the dynamical
status is also performed by using X--ray data stored in the Chandra
archive.}{We compute the line--of--sight (LOS) velocity dispersion of
galaxies, $\sigma_{\rm v}=1334_{-95}^{+114}$ \kss, in agreement with
the high average X--ray temperature $T_{\rm X}=$(10.2$\pm$0.8) keV
recovered from Chandra data, as expected in the case of
energy--density equipartition between galaxies and gas.  Assuming that
the cluster is in dynamical equilibrium and mass follows the galaxy
distribution, we find that A697 is a very massive cluster obtaining
$M(<{\rm R}_{\rm max}=0.75$ \hh)$=9.5^{+1.8}_{-1.5}$ \mqua and
$M(<{\rm R}_{\rm vir}=3.85$ \hh)$=4.5_{-1.3}^{+1.4}$ \mqui for the
region well sampled by the spectroscopic data and for the entire
virialized region, respectively.  Further investigations find that
A697 is not fully relaxed, as shown by the non Gaussianity of the
velocity distribution, the elongation of the X--ray emission, and the
presence of small-size substructures in the central region.}  {Our
results suggest that we are looking at a cluster undergone to a
complex cluster merger occurring roughly mainly along the LOS, with a
transverse component in the SSE--NNW direction. The importance and the
phase of the merging event is discussed. The spatial correlation
between the (likely) radio halo and the optical and X--ray cluster
structure supports the hypothesis of a relation between extended radio
emission and merging phenomena.}


\authorrunning{Girardi et al.}
\titlerunning{Internal dynamics of A697} 
\maketitle
%

\section{Introduction}

Clusters of galaxies are by now recognized to be not simple relaxed
structures, but rather they are evolving via merging processes in a
hierarchical fashion from poor groups to rich clusters. Much progress
has been made in recent years in the observations of the signatures of
merging processes (see Feretti et al. \cite{fer02b} for a general
review). The presence of substructure, which is indicative of a
cluster in an early phase of the process of dynamical relaxation or of
secondary infall of clumps into already virialized clusters, occurs in
about $50\%$ of clusters as shown by optical and X--ray data (e.g.,
Geller \& Beers \cite{gel82}; Mohr et al. \cite{moh96}; Girardi et
al. \cite{gir97}; Kriessler \& Beers \cite{kri97}; Jones \& Forman
\cite{jon99}; Schuecker et al. \cite{sch01}; Burgett et
al. \cite{bur04}).

A new aspect of these investigations is the possible connection of
cluster mergers with the presence of extended, diffuse radio sources,
halos and relics. They are rare, large (up to $\sim1$ \hh), amorphous
cluster sources of uncertain origin and generally steep radio spectra
(Hanisch \cite{han82}; see also Giovannini \& Feretti \cite{gio02} for
a recent review).  They appear to be associated with very rich
clusters that have undergone recent mergers and thus it has been
suggested by various authors that cluster halos/relics are related to
recent merger activity (e.g., Tribble \cite{tri93}; Burns et
al. \cite{bur94}; Feretti \cite{fer99}).

The synchrotron radio emission of halos and relics demonstrates the
existence of large scale cluster magnetic fields, of the order of
0.1--1 $\mu$G, and of widespread relativistic particles of energy
density 10$^{-14}$ -- 10$^{-13}$ erg cm$^{-3}$.  The difficulty in
explaining radio--halos arises from the combination of their large
size and the short synchrotron lifetime of relativistic electrons.
The expected diffusion velocity of the electron population is on the
order of the Alfven speed ($\sim100$ \kss) making it difficult for the
electrons to diffuse over a megaparsec--scale region within their
radiative lifetime. Therefore, one needs a mechanism by which the
relativistic electron population can be transported over large
distances in a short time, or a mechanism by which the local electron
population is reaccelerated and the local magnetic fields are
amplified over an extended region.  The cluster--cluster merger can
potentially supply both mechanisms (e.g., Giovannini et
al. \cite{gio93}; Burns et al. \cite{bur94}; R\"ottgering et
al. \cite{rot94}; see also Feretti et al. \cite{fer02a}; Sarazin
\cite{sar02} for reviews). However, the question is still debated
since the diffuse radio sources are quite uncommon and only recently
we can study these phenomena on the basis of a sufficient statistics
($\sim30$ clusters up to $z\sim0.3$, e.g., Giovannini et
al. \cite{gio99}; see also Giovannini \& Feretti \cite{gio02}; Feretti
\cite{fer05}).

Growing evidence of the connection between diffuse radio emission and
cluster merging is based on X--ray data (e.g., B\"ohringer \&
Schuecker \cite{boh02}; Buote \cite{buo02}). Studies based on a large
number of clusters have found a significant relation between the radio
and the X--ray surface brightness (Govoni et al. \cite{gov01a},
\cite{gov01b}) and between the presence of radio--halos/relics and
irregular and bimodal X--ray surface brightness distribution
(Schuecker et al. \cite{sch01}).  New unprecedent insights into
merging processes in radio clusters are offered by Chandra and
XMM--Newton observations (e.g., Markevitch \& Vikhlinin \cite{mar01};
Markevitch et al. \cite{mar02}; Fujita et al. \cite{fuj04}; Henry et
al. \cite{hen04}; Kempner \& David \cite{kem04}).

Optical data are a powerful way to investigate the presence and the
dynamics of cluster mergers, too (e.g., Girardi \& Biviano \cite{gir02}).
The spatial and kinematical analysis of member galaxies allow us
to detect and measure the amount of substructure, to identify and
analyze possible pre--merging clumps or merger remnants.  This optical
information is really complementary to X--ray information since
galaxies and ICM react on different time scales during a merger (see
numerical simulations by Roettiger et al. \cite{roe97}).
Unfortunately, to date optical data is lacking or poorly exploited.
Sparse literature concerns some few individual clusters (e.g., Colless
\& Dunn \cite{col96}; G\'omez et al. \cite{gom00}; Barrena et
al. \cite{bar02}; Mercurio et al. \cite{mer03}; Boschin et
al. \cite{bos04}; Boschin et al. \cite{bos06}). In this context we
have conducted an intensive study of Abell 697 (hereafter A697) having
a probable diffuse radio emission located in the cluster center
(Kempner \& Sarazin \cite{kem01}).

A697 is one of the higher redshift clusters in the ACO catalog
($z\sim0.282$, Abell et al. \cite{abe89}). It is a fairly rich,
X--ray luminous, and hot cluster: Abell richness class =1 (Abell et
al. \cite{abe89}), $L_\mathrm{X}$(0.1--2.4 keV)$=16.30\times 10^{44}$
h$_{50}^{-2}$ erg\ s$^{-1}$ (Ebeling et al. \cite{ebe98}),
$T_\mathrm{X}\sim8$--11 keV (e.g., Metzger \& Ma
\cite{met00}--hereafter M00; White \cite{whi00}; Ota \& Mitsuda
\cite{ota04}; Bonamente et al. \cite{bon05}).  It shows an arclike
feature as revealed by Keck images (M00) and is one of the most
massive clusters analyzed by Dahle et al. (\cite{dah02}) through the
weak lensing analysis.  Observational signatures of the young
dynamical state of A697 come from Keck images of the dominant
galaxy, a cD showing a highly asymmetric halo and a (likely)
secondary nucleus (M00).  Both X--ray emission and gravitational
lensing data indicate an elongated mass distribution (M00; Dahle et al
\cite{dah02}; De Filippis et al. \cite{def05}) as expected in a merger
collision (e.g., Roettiger et al. \cite{roe96}).

To date few spectroscopic data are reported in the literature.
Crawford et al. (\cite{cra95}) measured the redshift for the cD
galaxy. M00 measured redshift for other 7(9) member galaxies giving an
uncertain value of 553(941) \ks for the line--of--sight (hereafter
LOS) velocity dispersion. Recently, we have carried out spectroscopic
observations at the TNG telescope giving new redshift data for 93
galaxies in the field of A697, as well as photometric observations at
the INT telescope. Our present analysis is based on these optical data
and X--ray Chandra archival data.

This paper is organized as follows.  We present the new optical data
in Sect.~2 and the relevant analyses in Sects.~3 and 4.  Our analysis
of X--ray Chandra data is shown in Sect.~5.  We discuss the dynamical
state of A697 in Sect.~6 and summarize our results in Sect.~7.

Unless otherwise stated, we give errors at the 68\% confidence level
(hereafter c.l.).  Throughout the paper, we assume a flat cosmology
with $\Omega_{\rm m}=0.3$, $\Omega_{\Lambda}=0.7$, and $H_0=70$
$\mathrm{h_{70}}$ \ks Mpc$^{-1}$. For this cosmological model, 1\arcm
corresponds to 256 \kpc at the cluster redshift.

\section{Data sample}

Multi--object spectroscopic observations of A697 were carried out at
the TNG telescope in December 2003 during the program of proposal
AOT8/CAT-G6. We used DOLORES/MOS with the LR-B Grism 1, yielding a
dispersion of 187 \AA/mm, and the Loral CCD of $2048\times2048$ pixels
(pixel size of 15 $\mu$m). This combination of grating and detector
results in dispersions of 2.8 \AA/pix. We have taken three MOS masks
for a total of 114 slits. We acquired two exposures of 1800 s for two
masks and three exposures of 1800 s for the last one. Wavelength
calibration was performed using Helium-Argon lamps.

Reduction of spectroscopic data was carried out with IRAF
\footnote{IRAF is distributed by the National Optical Astronomy
Observatories, which are operated by the Association of Universities
for Research in Astronomy, Inc., under cooperative agreement with the
National Science Foundation.} package.

\begin{figure*}
\centering
\resizebox{\hsize}{!}{\includegraphics{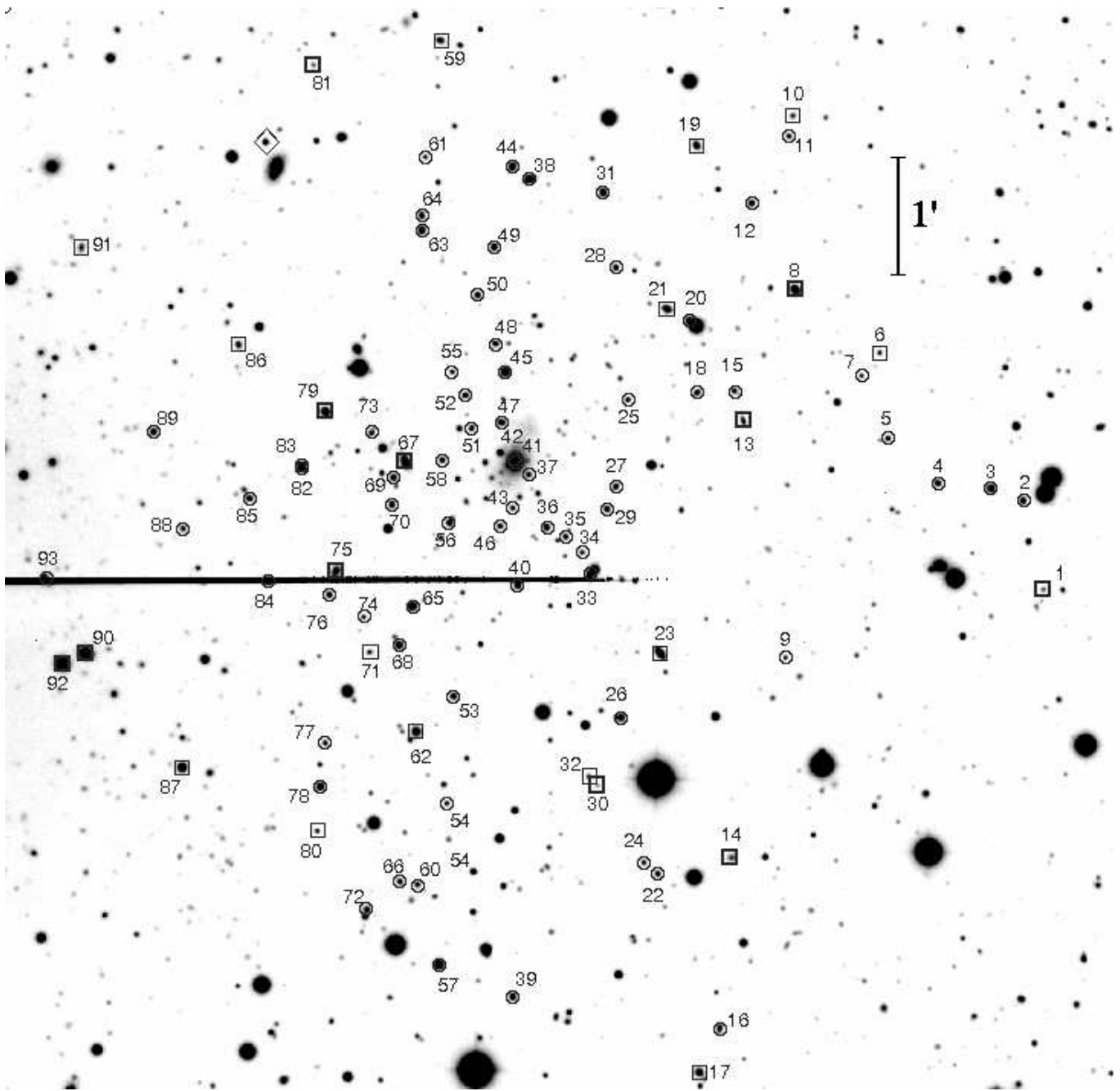}}
\caption{$R$--band image of A697 (North at the top and East to the
left) taken with the WFC camera of the INT. Galaxies with successful
velocity measurements are labeled as in Table~\ref{catalogue}. Circles
and boxes indicate cluster members and non member galaxies,
respectively. Out of non member galaxies, bold--face boxes indicate
emission--line galaxies. Diamond indicates a QSO at $z\sim1.50$. 
The horizontal spike is due to the presence of a bright star $\sim
7$ \arcm East of the cluster.}
\label{figimage}
\end{figure*}

\begin{figure*}[!ht]
\centering
\resizebox{\hsize}{!}{\includegraphics{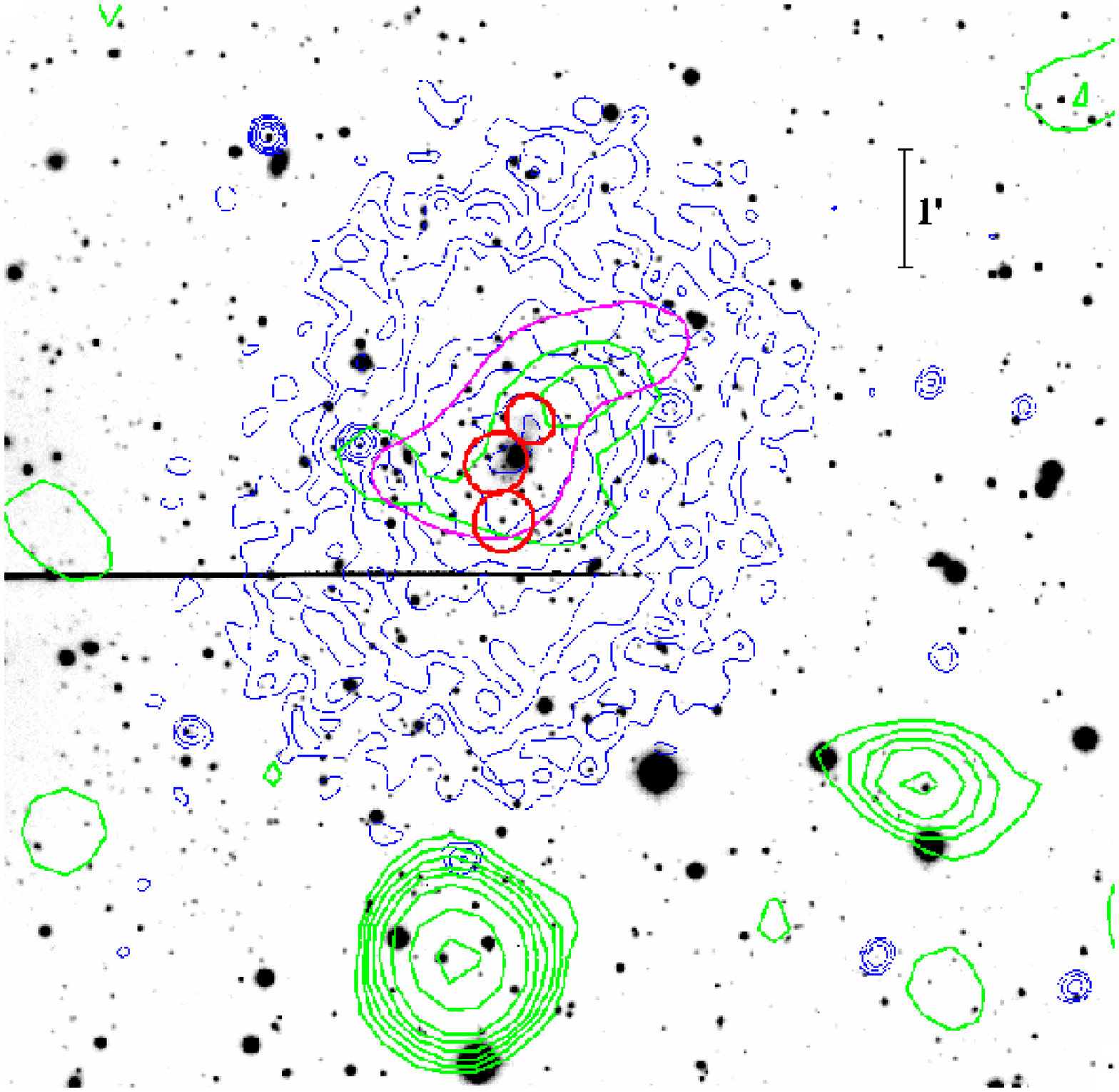}}
\caption{$R$--band image of the cluster A697 with, superimposed, the
contour levels of the Chandra X--ray image ID \#4217 (blue, photons in
the energy range 0.3--7 keV) and NVSS (Condon et al. \cite{con98})
radio image (green, see also Kempner \& Sarazin \cite{kem01}).  Red
ellipses identify structures detected by Wavdetect. To avoid
confusion, only one isodensity contour of the spatial distribution of
the (likely) cluster members is shown (magenta, see also
Fig.~\ref{fig2113}).  North is at the top and East to the left.}
\label{figisofote}
\end{figure*}

Radial velocities were determined using the cross--correlation
technique (Tonry \& Davis \cite{ton79}) implemented in the RVSAO
package (developed at the Smithsonian Astrophysical Observatory
Telescope Data Center).  Each spectrum was correlated against six
templates for a variety of galaxy spectral types: E, S0, Sa, Sb, Sc,
Ir (Kennicutt \cite{ken92}).  The template producing the highest value
of $\cal R$, i.e., the parameter given by RVSAO and related to the
signal--to--noise of the correlation peak, was chosen.  Moreover, all
spectra and their best correlation functions were examined visually to
verify the redshift determination.  In some cases (IDs~1, 30, 75, 81;
see Table~1) we took the EMSAO redshift, i.e. that determined
from the emission lines in the spectra, as a reliable
estimate of the redshift.  One object
[R.A.=$08^{\mathrm{h}}43^{\mathrm{m}}08\dotsec16$, Dec.=$+36\degree
24\arcmm 39\dotarcs1$ (J2000.0), see the diamond in
Fig.~\ref{figimage}] is found to be a quasar at $z\sim1.50$ (see also
the point--like X--ray emission in the upper--left corner of
Fig.~\ref{figisofote}) and has not been considered in our analysis.

For three galaxies we obtained two redshift determinations, which are
of similar quality.  We compared the two determinations computing the
mean and the rms of the variable $(z_1-z_2)/\sqrt{err_1^2+err_2^2}$,
where $z_1$ comes from MOS~1 and $z_2$ from MOS~2 (or MOS~3).  We
obtained a mean=0.23$\pm$0.43 and a rms$=0.74$, to be compared with
the expected values of 0 and 1.  The resulting mean shows that two
sets of measurements are consistent with having the same velocity
zero--point, and the value of rms is compatible with a value of 1
according to the $\chi^2$--test.  For the three galaxies we used the
average of the two redshift determinations and the corresponding
error.

Our spectroscopic catalog consists of 93 galaxies out to a radius of
R$\sim$ 5\arcm from the cD galaxy (ID~41).

\begin{table}
        \caption[]{Velocity catalog of 93 spectroscopically measured galaxies. In Column~1, IDs in italics indicate non--cluster galaxies.}
         \label{catalogue}
              $$ 
           \begin{array}{r c c c r r l}
            \hline
            \noalign{\smallskip}
            \hline
            \noalign{\smallskip}

\mathrm{ID} & \mathrm{\alpha},\mathrm{\delta}\,(\mathrm{J}2000)  & B& R  & \mathrm{v} & \mathrm{\Delta}\mathrm{v} & \mathrm{EL} \\
  & 08^{\rm h}      , +36^{\rm o}    &  &  &\,\,\,\,\mathrm{km}&\mathrm{s^{-1}}\,&\\
            \hline
            \noalign{\smallskip}   

  \textit{1}    &     42\ 35.42 , 20\ 58.0 & 21.77 & 20.76 & 109714 & 26 & \mathrm{[OII]}\\
  2             &     42\ 36.31 , 21\ 43.0 & 21.33 & 19.21 &  84766 & 91 &  \\
  3             &     42\ 37.68 , 21\ 49.0 & 21.05 & 18.77 &  84801 & 46 &  \\
  4             &     42\ 39.89 , 21\ 51.4 & 21.45 & 19.50 &  85868 & 54 &  \\
  5             &     42\ 42.00 , 22\ 13.7 & 22.26 & 19.94 &  84593 & 53 &  \\
  \textit{6}    &     42\ 42.38 , 22\ 56.9 & 22.51 & 20.25 &  87591 & 68 &  \\
  7             &     42\ 43.13 , 22\ 45.4 & 23.07 & 20.41 &  85957 & 82 &  \\
  \textit{8}    &     42\ 45.96 , 23\ 28.2 & 19.37 & 18.49 &   9076 & 63 & \mathrm{[OII],[OIII]}\\
  9             &     42\ 46.15 , 20\ 22.2 & 22.96 & 20.55 &  85453 & 95 &  \\
 \textit{10}    &     42\ 46.15 , 24\ 56.2 & 22.50 & 20.05 & 154872 & 43 &  \\
 11             &     42\ 46.32 , 24\ 45.8 & 22.22 & 19.95 &  83814 & 48 &  \\
 12             &     42\ 47.83 , 24\ 11.7 & 21.77 & 19.37 &  85195 & 52 &  \\
 \textit{13}    &     42\ 48.05 , 22\ 21.8 & 20.90 & 19.70 &  82551 &250 & \mathrm{[OII]}\\
 \textit{14}    &     42\ 48.34 , 18\ 40.8 & 20.64 & 19.69 &  15597 &127 & \mathrm{[OIII], H\alpha}\\
 15             &     42\ 48.38 , 22\ 36.4 & 21.57 & 19.67 &  81170 & 82 &  \\
 16             &     42\ 48.67 , 17\ 14.4 & 21.40 & 19.33 &  85836 & 65 &  \\
 \textit{17}    &     42\ 49.46 , 16\ 52.1 & 21.45 & 18.80 & 103561 & 89 &  \\
 18             &     42\ 49.97 , 22\ 35.7 & 21.97 & 19.59 &  86779 & 58 &  \\
 \textit{19}    &     42\ 50.16 , 24\ 40.6 & 21.43 & 19.07 &  87839 & 60 &  \\
 20             &     42\ 50.36 , 23\ 12.1 & 21.80 & 19.49 &  82694 & 75 &  \\
 \textit{21}    &     42\ 51.31 , 23\ 17.8 & 20.88 & 18.85 & 102192 & 56 &  \\
 22             &     42\ 51.36 , 18\ 32.2 & 22.21 & 19.70 &  84197 & 82 &  \\
 \textit{23}    &     42\ 51.38 , 20\ 23.9 & 20.06 & 18.25 &  41336 & 41 &  \\
 24             &     42\ 51.91 , 18\ 37.9 & 22.59 & 20.09 &  83253 & 63 &  \\
 25             &     42\ 52.87 , 22\ 31.8 & 22.48 & 20.06 &  83404 & 68 &  \\
 26             &     42\ 52.99 , 19\ 51.0 & 20.37 & 18.81 &  86695 & 49 &  \\
 27             &     42\ 53.33 , 21\ 47.8 & 22.23 & 19.59 &  85917 & 75 &  \\
 28             &     42\ 53.47 , 23\ 38.5 & 22.16 & 19.73 &  82697 & 77 &  \\
 29             &     42\ 53.69 , 21\ 36.3 & 21.76 & 19.42 &  87319 & 74 &  \\
 \textit{30}    &     42\ 53.93 , 19\ 16.6 & 22.24 & 21.05 & 101857 &311 & \mathrm{[OII],[OIII]}\\
 31             &     42\ 54.05 , 24\ 16.5 & 21.42 & 18.86 &  85853 & 43 &  \\
 \textit{32}    &     42\ 54.26 , 19\ 21.2 & 22.46 & 20.30 & 101888 & 57 &  \\
 33             &     42\ 54.36 , 21\ 03.5 & 21.37 & 19.83 &  82240 & 53 &  \\
 34             &     42\ 54.67 , 21\ 14.2 & 21.94 & 19.99 &  85332 & 43 &  \\
 35             &     42\ 55.39 , 21\ 22.1 & 22.30 & 19.50 &  82513 & 63 &  \\
              \noalign{\smallskip}			    
            \hline					    
            \noalign{\smallskip}			    
            \hline					    
         \end{array}
     $$ 
         \end{table}
\addtocounter{table}{-1}
\begin{table}
          \caption[ ]{Continued.}
     $$ 
           \begin{array}{r c c c r r l}
            \hline
            \noalign{\smallskip}
            \hline
            \noalign{\smallskip}

\mathrm{ID} & \mathrm{\alpha},\mathrm{\delta}\,(\mathrm{J}2000)  & B & R  & \mathrm{v} & \mathrm{\Delta}\mathrm{v} & \mathrm{EL} \\
  & 08^{\rm h}      , +36^{\rm o}    &  &  &\,\,\,\,\,\mathrm{km}&\mathrm{s^{-1}}\,&\\
            \hline
            \noalign{\smallskip}
 36             &     42\ 56.18 , 21\ 26.7 & 21.90 & 19.04 &  85475 & 41 &  \\
 37             &     42\ 56.98 , 21\ 53.3 & 22.00 & 19.82 &  85011 & 97 &  \\
 38             &     42\ 57.12 , 24\ 22.6 & 20.47 & 18.47 &  87199 & 33 &  \\
 39             &     42\ 57.31 , 17\ 29.4 & 21.50 & 19.22 &  82900 & 66 &  \\
 40             &     42\ 57.38 , 20\ 57.4 & 21.50 & 19.43 &  87821 & 49 &  \\
 41             &     42\ 57.55 , 21\ 59.9 & 20.48 & 17.95 &  84361 & 58 &  \\
 42             &     42\ 57.60 , 22\ 01.2 & 21.90 & 19.74 &  84711 & 45 &  \\
 43             &     42\ 57.62 , 21\ 36.5 & 22.18 & 19.49 &  82422 & 67 &  \\
 44             &     42\ 57.84 , 24\ 29.0 & 21.16 & 18.77 &  84988 & 43 &  \\
 45             &     42\ 58.01 , 22\ 45.3 & 20.95 & 18.30 &  83468 & 48 &  \\
 46             &     42\ 58.13 , 21\ 27.3 & 21.66 & 19.41 &  85455 & 85 &  \\
 47             &     42\ 58.13 , 22\ 19.5 & 21.06 & 18.65 &  84221 & 37 &  \\
 48             &     42\ 58.44 , 22\ 59.1 & 22.35 & 19.58 &  87125 & 57 &  \\
 49             &     42\ 58.56 , 23\ 48.2 & 21.59 & 19.18 &  81888 & 77 &  \\
 50             &     42\ 59.21 , 23\ 23.8 & 21.83 & 19.53 &  82610 & 67 &  \\
 51             &     42\ 59.38 , 22\ 16.5 & 22.20 & 19.47 &  86794 & 56 &  \\
 52             &     42\ 59.66 , 22\ 33.0 & 22.30 & 19.55 &  84290 &130 &  \\
 53             &     42\ 59.98 , 20\ 00.9 & 21.85 & 19.44 &  86525 & 46 &  \\
 54             &     43\ 00.19 , 19\ 06.5 & 22.68 & 20.34 &  82625 & 73 &  \\
 55             &     43\ 00.24 , 22\ 44.9 & 22.63 & 20.22 &  83829 & 98 &  \\
 56             &     43\ 00.29 , 21\ 28.4 & 21.98 & 19.62 &  86920 & 52 &  \\ 
 57             &     43\ 00.43 , 17\ 45.1 & 20.64 & 18.20 &  84509 & 48 &  \\
 58             &     43\ 00.58 , 22\ 00.2 & 22.39 & 19.73 &  86018 & 71 &  \\
 \textit{59}    &     43\ 00.91 , 25\ 32.3 & 21.11 & 19.30 &  89555 & 75 &  \\
 60             &     43\ 01.34 , 18\ 24.9 & 22.12 & 19.66 &  84617 & 61 &  \\
 61             &     43\ 01.49 , 24\ 33.2 & 22.79 & 20.37 &  82420 & 55 &  \\
 \textit{62}    &     43\ 01.51 , 19\ 42.7 & 20.74 & 18.49 &  80752 & 39 &  \\
 63             &     43\ 01.58 , 23\ 56.3 & 21.56 & 19.10 &  82712 & 51 &  \\
 64             &     43\ 01.61 , 24\ 03.7 & 21.68 & 19.44 &  83882 & 59 &  \\
 65             &     43\ 01.73 , 20\ 46.1 & 21.06 & 18.79 &  85345 & 36 &  \\
 66             &     43\ 02.09 , 18\ 26.7 & 22.09 & 19.83 &  84853 & 79 &  \\
 \textit{67}    &     43\ 02.14 , 21\ 59.1 & 19.95 & 18.05 &  79922 & 37 & \mathrm{[OII],[OIII]}\\
 68             &     43\ 02.28 , 20\ 26.4 & 21.29 & 18.92 &  83675 & 63 &  \\
 69             &     43\ 02.62 , 21\ 51.0 & 21.93 & 19.42 &  82818 & 69 &  \\
 70             &     43\ 02.66 , 21\ 37.4 & 21.78 & 19.50 &  85424 & 27 &  \\
              \noalign{\smallskip}			    
            \hline					    
            \noalign{\smallskip}			    
            \hline					    
         \end{array}
     $$ 
         \end{table}
\addtocounter{table}{-1}
\begin{table}
          \caption[ ]{Continued.}
     $$ 
           \begin{array}{r c c c r r l}
            \hline
            \noalign{\smallskip}
            \hline
            \noalign{\smallskip}

\mathrm{ID} & \mathrm{\alpha},\mathrm{\delta}\,(\mathrm{J}2000)  & B & R  & \mathrm{v} & \mathrm{\Delta}\mathrm{v} & \mathrm{EL} \\
  & 08^{\rm h}      , +36^{\rm o}    &  &  &\,\,\,\,\,\mathrm{km}&\mathrm{s^{-1}}\,&\\
            \hline
            \noalign{\smallskip}
 \textit{71}    &     43\ 03.46 , 20\ 22.6 & 22.43 & 20.13 & 110946 & 98 &  \\
 72             &     43\ 03.48 , 18\ 12.6 & 22.29 & 19.72 &  84978 & 49 &  \\
 73             &     43\ 03.55 , 22\ 14.0 & 21.22 & 19.75 &  85604 & 57 &  \\
 74             &     43\ 03.74 , 20\ 40.6 & 22.48 & 20.29 &  80695 & 83 &  \\
 \textit{75}    &     43\ 04.94 , 21\ 03.3 & 20.97 & 19.78 &  15299 & 85 & \mathrm{[OII],[OIII],[OI],H\alpha} \\
 76             &     43\ 05.23 , 20\ 51.6 & 21.78 & 19.63 &  82393 & 63 &  \\
 77             &     43\ 05.30 , 19\ 36.4 & 22.77 & 20.07 &  83863 & 57 &  \\
 78             &     43\ 05.47 , 19\ 14.4 & 21.48 & 18.96 &  83856 & 58 &  \\
 \textit{79}    &     43\ 05.50 , 22\ 23.9 & 19.89 & 18.70 &  80047 & 35 & \mathrm{[OII],[OIII]}\\
 \textit{80}    &     43\ 05.57 , 18\ 51.9 & 22.87 & 20.57 &  90606 &106 &  \\
 \textit{81}    &     43\ 06.22 , 25\ 19.1 & 22.25 & 20.66 &  77636 & 71 & \mathrm{[OII]}\\
 82             &     43\ 06.45 , 21\ 55.0 & 21.23 & 18.88 &  82376 & 44 &  \\
 83             &     43\ 06.46 , 21\ 56.8 & 22.72 & 20.32 &  84841 & 67 &  \\
 84             &     43\ 07.80 , 20\ 58.1 & 21.41 &  ...  &  85932 & 29 &  \\
 85             &     43\ 08.62 , 21\ 39.8 & 21.56 & 19.43 &  81414 & 54 &  \\
 \textit{86}    &     43\ 09.17 , 22\ 57.2 & 22.24 & 19.67 &  87994 &100 &  \\
 \textit{87}    &     43\ 11.23 , 19\ 23.3 & 20.60 & 18.64 &  87555 & 52 &  \\
 88             &     43\ 11.38 , 21\ 24.0 & 22.54 & 20.11 &  85718 &134 &  \\
 89             &     43\ 12.67 , 22\ 12.9 & 21.84 & 19.45 &  83517 & 54 &  \\
 \textit{90}    &     43\ 15.34 , 20\ 20.2 & 18.87 & 17.71 &  32968 & 39 & \mathrm{[OII],H\beta,[OIII],H\alpha,[SII]}\\
 \textit{91}    &     43\ 15.79 , 23\ 45.5 & 21.51 & 19.84 & 104134 & 83 &  \\
 \textit{92}    &     43\ 16.30 , 20\ 14.9 & 18.92 & 17.42 &  25661 & 41 & \mathrm{H\alpha,[SII]}\\
 93             &     43\ 17.04 , 20\ 58.4 & 21.46 & 20.12 &  85899 & 86 &  \\
              \noalign{\smallskip}			    
            \hline					    
            \noalign{\smallskip}			    
            \hline					    
         \end{array}\\
     $$ 
\\
   \end{table}

As far as photometry is concerned, our observations were carried out
with the Wide Field Camera (WFC), mounted at the prime focus of the
2.5m INT telescope (located at Roque de los Muchachos Observatory, La
Palma, Spain). We observed A697 on December 18th 2004 under
photometric conditions with a seeing of about 2\arcs.

The WFC consists of a 4 CCDs mosaic covering a 30\arcmm$\times$30\arcm
field of view, with only a $20\%$ marginally vignetted area. We took
10 exposures of 720 s in $B_{\rm H}$ and 360 s in $R_{\rm H}$ Harris
filters (a total of 7200 s and 3600 s in each band) developing a
dithering pattern of ten positions. This observing mode allowed us to
build a ``supersky'' frame that was used to correct our images for
fringing patterns (Gullixson \cite{gul92}). In addition, the dithering
helped us to clean cosmic rays and avoid gaps between the CCDs in the
final images. The complete reduction process (including flat fielding,
bias subtraction and bad-columns elimination) yielded a final coadded
image where the variation of the sky was lower than 0.4$\%$ in the
whole frame.  Another effect associated with the wide field frames is
the distortion of the field. In order to match the photometry of
several filters, a good astrometric solution is needed to take into
account these distortions. Using IRAF tasks and taking as reference
the USNO B1.0 catalog we were able to find an accurate astrometric
solution (rms $\sim$ 0.5\arcss) across the full frame. The photometric
calibration was performed using Landolt standard fields achieved
during the observation.

We finally identified galaxies in our $B_{\rm H}$ and $R_{\rm H}$
images and measured their magnitudes with the SExtractor package
(Bertin \& Arnouts \cite{ber96}) and AUTOMAG procedure. In few cases,
(e.g., close companion galaxies, galaxies close to defects of CCD),
the standard SExtractor photometric procedure failed. In these cases
we computed magnitudes by hand. This method consists in assuming a
galaxy profile of a typical elliptical and scaling it to the maximum
observed value. The integration of this profile gives us an estimate of
the magnitude. The idea of this method is similar to the PSF
photometry, but assuming a galaxy profile, more appropriate in this
case.

We transformed all magnitudes into the Johnson-Cousins system (Johnson
\& Morgan \cite{joh53}; Cousins \cite{cou76}). We used
$B=B\rm_H+0.13$ and $R=R\rm_H$, as derived from the Harris filter
characterization
(http://www.ast.cam.ac.uk/$\sim$wfcsur/technical/photom/colours/) and
assuming a $B$--$V\sim1.0$ for E-type galaxies (Poggianti
\cite{pog97}).  As a final step, we estimated and corrected the
galactic extinction $A_B \sim0.15$, $A_R \sim0.09$ from Burstein \&
Heiles (\cite{bur82}) reddening maps.

We estimated that our photometric sample is complete down to $R=22.0$
(23.2) and $B=23.0$ (24.2) for $S/N=5$ (3) within the observed field.

We assigned $R$ ($B$) magnitudes to 92 (93) out of the 93 galaxies of
our spectroscopic catalog.  We measured redshifts for galaxies down to
magnitude $R\sim$ 21, but a high level of completeness is reached
only for galaxies with magnitude $R<$ 20 ($\sim$70\% completeness).

Table~\ref{catalogue} lists the velocity catalog (see also
Fig.~\ref{figimage}): identification number of each galaxy, ID
(Col.~1); right ascension and declination, $\alpha$ and $\delta$
(J2000, Col.~2); $B$ magnitudes (Col.~3); $R$ magnitudes (Col.~4);
heliocentric radial velocities, ${\rm v}=cz_{\sun}$ (Col.~5) with
errors, $\Delta {\rm v}$ (Col.~6); emission lines detected in the
spectra (Col.~7).

\section{Analysis of the spectroscopic catalog}

\subsection{Member selection}

Out of 93 galaxies having redshifts, the identification of cluster
members proceeds in two steps, following a procedure already used for
nearby and medium--redshift clusters (Fadda et al. \cite{fad96};
Girardi et al. \cite{gir96}; Girardi \& Mezzetti \cite{gir01}).

\begin{figure}
\centering
\resizebox{\hsize}{!}{\includegraphics{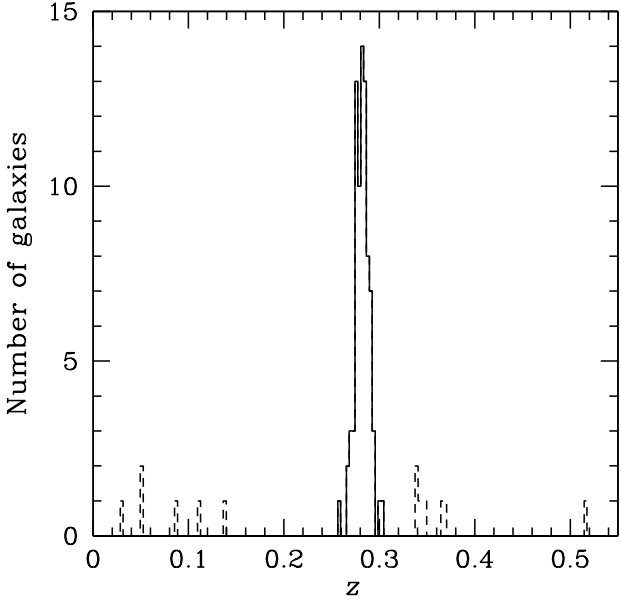}}
\caption
{Redshift galaxy distribution. The solid--line histogram 
refers to galaxies assigned to the cluster peak according to
the adaptive--kernel reconstruction method.}
\label{figden}
\end{figure}

First, we perform the cluster--member selection in velocity space by
using only redshift information. We apply the adaptive--kernel method
(Pisani \cite{pis93}) to find the significant ($>99\%$ c.l.)  peaks in
the velocity distribution.  This procedure detects A697 as a well
isolated peak at $\left<z\right>=0.282$ assigning 79 galaxies
considered as candidate cluster members (see Fig.~\ref{figden}).  Out
of non--member galaxies, six and eight are foreground and background
galaxies, respectively.

\begin{figure}
\centering 
\resizebox{\hsize}{!}{\includegraphics{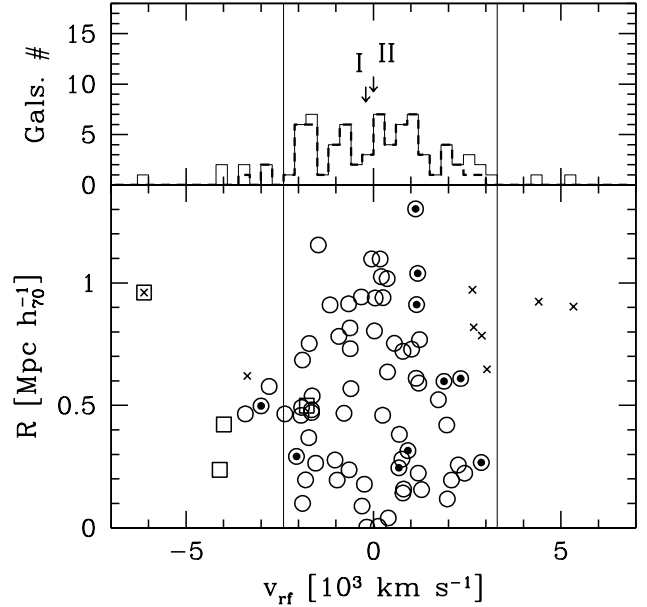}}
\caption
{ {\em Lower panel}: rest--frame velocity vs. projected clustercentric
distance for the 79 galaxies in the main peak (Fig.~\ref{figden}),
where we indicate galaxies rejected by the ``shifting gapper'' method
(crosses) and emission line galaxies (squares).  The remaining 68
fiducial cluster--members (sample--A) are indicated by open circles
(out of which blue galaxies are indicated by a small solid circle, see
Sect.~3.3).  {\em Upper panel}: velocity distribution of all 79
galaxies in the main peak (faint solid--line) and 68 member galaxies
(dashed line).  Velocities of the main and secondary nuclei of the cD
galaxy are pointed out (IDs~41 and 42, respectively).  In both panels
faint vertical lines indicate velocity boundaries of the alternative
sample of member galaxies (sample--B, see text).  }
\label{figvd}
\end{figure}

All the galaxies assigned to the A697 peak are analyzed in the second
step, which uses the combination of position and velocity information.
We apply the procedure of the ``shifting gapper'' by Fadda et
al. (\cite{fad96}).  This procedure rejects galaxies that are too far
in velocity from the main body of galaxies and within a fixed bin that
shifts along the distance from the cluster center.  The procedure is
iterated until the number of cluster members converges to a stable
value.  Following Fadda et al. (\cite{fad96}) we use a gap of $1000$
\ks in the cluster rest--frame and a bin of 0.6 \hh, or large enough
to include 15 galaxies.  As for the cluster center, we consider the
position of the cD galaxy
[R.A.=$08^{\mathrm{h}}42^{\mathrm{m}}57\dotsec55$, Dec.=$+36\degree
21\arcmm 59\dotarcs9$ (J2000.0)].  The shifting--gapper procedure
rejects eight galaxies as non--members (cross symbols in
Fig.~\ref{figvd}).  Following Girardi \& Mezzetti (\cite{gir01}) we
also reject three emission line galaxies: two of them dist $\sim4000$
\ks from the mean velocity in the cluster rest frame, the third has an
uncertain redshift since determined by only one emission line.

The member selection procedure leads to a sample of 68 cluster members
(see Table~\ref{catalogue}, Fig.~\ref{figimage}, and
Fig.~\ref{figvd}).

The above member selection was planned to be suitable for a single
galaxy system. Since the inspection of the velocity distribution of
A697 (see also Sect.~3.4) suggests the possible presence of many
structures, we also consider the following alternative member
selection.  We select all (non emission--line) galaxies in the range
v$\in$[81888 \kss, 87994 \kss], thus to consider all the apparent
overdensities in the velocity distribution and reject the
underpopulated tails. We obtain a sample of 69 galaxies.  When
necessary to avoid confusion, we refer to this alternative sample as
sample--B, and to the main one as sample--A.  The superposition between
the two samples is very large concerning 65 galaxies.  If not
explicitly said, our analyses always refers to the main sample--A.

\subsection{Global properties}

By applying the biweight estimator to cluster members (Beers et
al. \cite{bee90}), we compute a mean cluster redshift of
$\left<z\right>=0.2815\pm$ 0.0005, i.e.
$\left<\rm{v}\right>=(84505\pm$163) \kss.  We estimate the LOS
velocity dispersion, $\sigma_{\rm v}$, by using the biweight estimator
and applying the cosmological correction and the standard correction
for velocity errors (Danese et al. \cite{dan80}).  We obtain
$\sigma_{\rm v}=1334_{-95}^{+114}$ \kss, where errors are estimated
through a bootstrap technique.

\begin{figure}
\centering
\resizebox{\hsize}{!}{\includegraphics{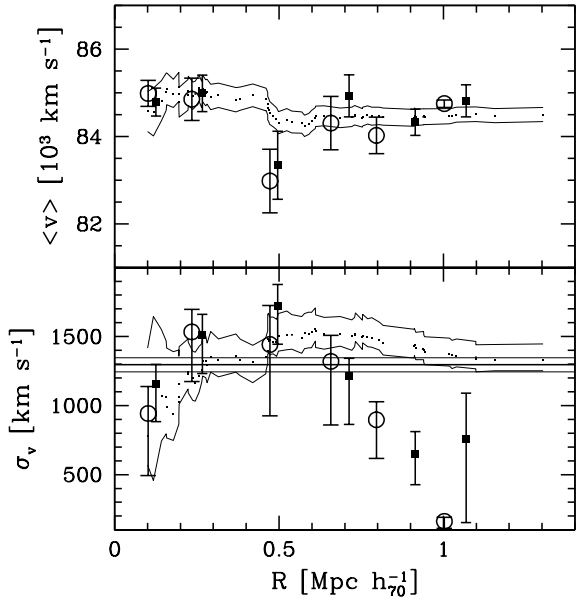}}
\caption
{Differential (solid squares) and integral (small points) profiles of
mean velocity ({\em upper panel}) and LOS velocity dispersion ({\em
lower panel}).  As for the differential profiles, results for six
annuli from the cluster center, each of 0.2 \hh, are shown.  As for
the integral profiles, the mean and dispersion at a given (projected)
radius from the cluster center is estimated by considering all
galaxies within that radius -- the first point is obtained on the
basis of the five galaxies close to the cluster center. In the lower
panel, the horizontal line represents the X--ray temperature with the
respective 68 per cent errors transformed in $\sigma_{\rm v}$ assuming
the density--energy equipartition between gas and galaxies, i.e.
$\beta_{\rm spec}=1$ (see text). In both panels differential profiles
for red galaxies is also shown (open circles).}
\label{figprof}
\end{figure}

To evaluate the robustness of the $\sigma_{\rm v}$ estimate we analyze
the velocity dispersion profile (see Fig.~\ref{figprof}).  The
integral profile smoothly decreases and then flattens in external
cluster regions (at $\sim1$ \hh) suggesting that a robust value of
$\sigma_{\rm v}$ is asymptotically reached, as found for most nearby
clusters (e.g., Fadda et al. \cite{fad96}; Girardi et
al. \cite{gir96}).  Both integral and differential velocity dispersion
profiles sharply rise in the central region out to $\sim0.5$ \hh.
This behaviour might be a signature of a relaxed cluster due to
circular velocities and galaxy mergers phenomena in the central
cluster region (e.g., Merritt \cite{mer87}; Merritt \cite{mer88};
Menci \& Fusco Femiano \cite{men96}; Girardi et al. \cite{gir98}) or,
alternatively, the consequence of the presence of subclumps, having
different mean velocities, LOS aligned in the central region.

In the next sections we further analyze the internal structure of A697
following two alternative possibilities: a) the presence of velocity
and spatial segregation of galaxies with respect to their colours and
luminosities, which is often taken as evidence of advanced dynamical
evolution of the parent cluster (Sect.~3.3); b) the presence of
substructures which is indicative of a cluster still far from a
complete dynamical relaxation (Sects.~3.4 and 3.5).

Here we assume that A697 is in dynamical equilibrium to compute virial
global quantities.  Following the prescriptions of Girardi \& Mezzetti
(\cite{gir01}), we assume for the radius of the quasi--virialized
region R$_{\rm vir}=0.17\times \sigma_{\rm v}/H(z) = 3.85$ \h -- see
their eq.~1 after introducing the scaling with $H(z)$ (see also eq.~ 8
of Carlberg et al. \cite{car97} for R$_{200}$). Therefore, we have
redshifts for galaxies out to a radius of R$_{\rm
out}\sim5$\arcmm$\sim0.34\times \rm{R}_{\rm vir}$ from the cluster center
and we well sample the region within R$_{\rm
max}\sim3$\arcmm$\sim0.2\times \rm{R}_{\rm vir}$.

One can compute the mass using the virial theorem (Limber \& Mathews
\cite{lim60}; see also, e.g., Girardi et al. \cite{gir98}) under the
assumption that mass follows galaxy distribution and using the data
for the $\rm{N}_{\rm g}$ observed galaxies:

\begin{equation}
M=M_{\rm svir}-\rm{SPT},
\end{equation}

\noindent where 
\begin{equation}
M_{\rm svir}=3\pi/2 \cdot \sigma_{\rm v}^2\rm{R}_{\rm PV}/G
\end{equation}

\noindent is the standard virial mass and SPT is the surface pressure
term correction (The \& White \cite{the86}). The size R$_{\rm PV}$, equal
to two times the (projected) harmonic radius, is:

\begin{equation}
\rm{R}_{\rm PV}=\rm{N}_{\rm g}(\rm{N}_{\rm g-1})/(\Sigma_{i>j}\rm{R}_{ij}^{-1}),
\end{equation}

\noindent 
where R$_{\rm ij}$ is the projected distance between two galaxies.

The estimate of $\sigma_{\rm v}$ is generally robust when computed
within a large cluster region (see Fig.~\ref{figprof} for A697 and
Fadda et al.  \cite{fad96} for other examples) and thus we consider
our global value.  The value of R$_{\rm PV}$ depends on the size of
the sampled region and possibly on the quality of the spatial sampling
(e.g., whether the cluster is uniformly sampled or not).  Here we
consider the well sampled region within R$_{\rm max}$ obtaining
R$_{\rm PV}=(0.75\pm$0.06) \hh, where the error is obtained via a
jacknife procedure.  The value of SPT correction strongly depends on
the amount of the radial component of the velocity dispersion at the
radius of the considered region and could be obtained by analyzing the
velocity--dispersion profile, although this procedure would require
several hundreds of galaxies.  Combining data on many clusters one
obtains that velocities are isotropic and that the SPT correction at
$\sim$R$_{\rm vir}$ is SPT$=0.2\cdot M_{\rm svir}$ (e.g., Carlberg
et al. \cite{car97}; Girardi et al. \cite{gir98}). Due to the limited
extension of our sample we prefer to recompute the SPT correction using
the eq.~8 of Girardi et al.~(\cite{gir98}) and assuming a galaxy
King--like distribution with parameters typical of
nearby/medium--redshift clusters: a core radius R$_{\rm c}=1/20\times
{\rm R}_{\rm vir}$ and a slope--parameter $\beta_{\rm fit}=0.8$,
i.e. the volume galaxy density at large radii goes as $r^{-3
\beta_{\rm fit}}=r^{-2.4}$ (Girardi \& Mezzetti \cite{gir01}).  We
obtain SPT$=0.35\cdot M_{\rm svir}$ and $M(<{\rm R}_{\rm max}=0.75\,
\hhh)=9.5^{+1.8}_{-1.5}$ \mquaa.

Calling into question the quality of the spatial sampling, one could
use an alternative estimate of R$_{\rm PV}$ on the basis of the
knowledge of the galaxy distribution (see eq.~13 of Girardi et
al. \cite{gir98}).  We obtain R$_{\rm PV}=0.73$ \h where a $25\%$ of
error is expected due to the fact that typical, rather than
individual, galaxy distribution parameters are assumed.  This leads to
a mass of $M(<{\rm R}_{\rm max}=0.75\,\hhh)=9.3^{+2.8}_{-2.7}$ \mquaa, in
good agreement with our first estimate.

We can use the second of the above approaches to obtain the mass
within the whole assumed virialized region, which is larger than that
sampled by observations $M(<{\rm R}_{\rm vir}=3.85\,
\hhh)=4.5_{-1.3}^{+1.4}$ \mquii, where we use a $20\%$
SPT--correction, well suitable in this case.

\subsection{Testing galaxy segregation}

The presence of velocity and spatial segregation of galaxies with
respect to their colours, luminosities, and morphologies is often
taken as evidence of advanced dynamical evolution of the parent
cluster (e.g., Biviano et al. \cite{biv92}; Ellingson et
al. \cite{ell01}; Biviano et al. \cite{biv02}; Poggianti \cite{pog04};
Goto \cite{got05}).  Here we check for possible luminosity and colour
segregation of galaxies, both in position and in velocity space, by
using the sample of 67 galaxies to which we have successfully assigned
$R$ magnitude and $B$--$R$ colours out of the 68 member galaxies.

As for luminosity segregation, we find no significant correlation
between the absolute velocity $|{\rm v}|$ and $R$ magnitude, as well
as between clustercentric distance R and $R$ magnitude.  We also
divide the sample into low-- and high--luminosity subsamples by using
the median $R$ magnitude=19.5.  To test for different means and
variances in velocity distributions of less-- and more--luminous
galaxies, we apply the standard means--test and F--test (e.g., Press
et al. \cite{pre92}).  We obtain no significant difference.  Moreover,
we verify that the two subsamples do not have different distributions
of v, R, 2D spatial positions, and [v,R] using the 1D and 2D
Kolmogorov--Smirnov tests (hereafter 1DKS and 2DKS--tests. See, e.g.,
Ledermann \cite{led82} for 1DKS; see Fasano \& Franceschini
\cite{fas87} for 2DKS, as implemented by Press et al. \cite{pre92}).

\begin{figure}
\centering
\resizebox{\hsize}{!}{\includegraphics{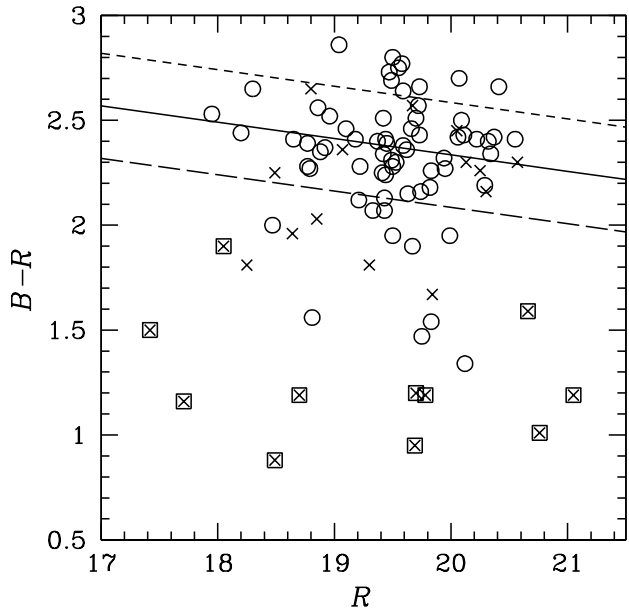}}
\caption
{$B$--$R$ vs. $R$ diagram for galaxies with available spectroscopy: circles
and crosses denote cluster and field members, respectively.  Out of
field members, squares denote emission line galaxies.  
The solid line gives the best--fit
colour--magnitude relation; the dashed lines are drawn at $\pm$0.25
mag from the CMR. According to our working definition in Sect.~3.3
cluster members are divided in: blue and red galaxies (below and above
the long--dashed line, respectively). In particular, very/not very red
galaxies lie above/down the solid line.}
\label{figcm}
\end{figure}

As for colour segregation, we find only a very marginal
anti--correlation between the clustercentric distance R and $B$--$R$
colour (at the $93.27\%$ c.l., according to the Kendall
non--parametric rank correlation test).  We also use the
colour--magnitude relation (hereafter CMR), which indicates the
early--type galaxy locus, to divide the sample into some subsets and
then compare their properties. To determine CMR we fix the slope
according to L\'opez--Cruz et al. (\cite{lop04}, see their Fig.~3) and
apply the two--sigma--clipping fitting procedure to the cluster
members obtaining $B$--$R=3.895-0.0780\times R$ (see
Fig.~\ref{figcm}).  ``Blue'' objects are defined to be those galaxies
at least 0.25 mag bluer in $B$--$R$ than the colour of the CMR. These
blue galaxies have typically $B$--$R\lesssim2$, thus include spiral
galaxies and exclude elliptical galaxies, according to the expected
typical colours at the cluster redshift (Buta et al. \cite{but94};
Buta \& Williams \cite{but95}; Poggianti \cite{pog97}).  Thus, the
remaining objects, which we define the ``red'' sample, should consist
mostly of ellipticals and lenticulars.  We also divide red galaxies in
``very red'' and ``not very red'' depending if they lie above or below
the CMR.  The blue sample (marginally) differs from the red sample as
concerning the 2D galaxy--position distribution, the distribution in
the [v,R] plane, and the velocity distribution (at the $93.40\%$,
$94.20\%$, and $94.45\%$ c.ls., according to the KS--tests,
respectively).  Instead, no difference is found between very red and
not very red galaxies.

\begin{table}
        \caption[]{Results of Kinematical analysis}
         \label{tabv}
                $$
         \begin{array}{l r l l}
            \hline
            \noalign{\smallskip}
            \hline
            \noalign{\smallskip}
\mathrm{Sample} & \mathrm{N_g} & \phantom{249}\mathrm{<v>}\phantom{249} & 
\phantom{24}\sigma_{\rm v}^{\mathrm{a}}\phantom{24}\\
& &\phantom{249}\mathrm{km\ s^{-1}}\phantom{249} 
&\phantom{2}\mathrm{km\ s^{-1}}\phantom{24}\\
            \hline
            \noalign{\smallskip}
 
\mathrm{Whole\ system\ (sample\ A)} & 68 &84505\pm163 &1334_{-95}^{+114}\\
\mathrm{Blue\ galaxies }         & 11 &85954\pm487 &1509_{-507}^{+763}\\
\mathrm{Red\ galaxies }          & 56 &84288\pm167 &1243_{-85}^{+120}\\
\mathrm{Whole\ system\ (sample\  B)} & 69 &84790\pm163 &1349_{-63}^{+112}\\
\mathrm{Sample\ A\ without\ KMM4g3\ gals} & 61 &84296\pm174 &1353_{-97}^{+122}\\
              \noalign{\smallskip}
            \hline
            \noalign{\smallskip}
            \hline
         \end{array}
$$
\begin{list}{}{}  
\item[$^{\mathrm{a}}$] We use the biweigth and the gapper estimators by
Beers et al. (1990) for samples with $\mathrm{N_g}\ge$ 15 and with
$\mathrm{N_g}<15$ galaxies, respectively (see also Girardi et
al. \cite{gir93}).
\end{list}
         \end{table}

Table~\ref{tabv} shows $\left<{\rm v}\right>$ and $\sigma_{\rm v}$
estimates for blue and red galaxies: the $\left<{\rm v}\right>$ of
blue galaxies is higher than that of red galaxies at the $97.37\%$
c.l., according to the means--test. The difference we detect in the
mean velocity suggests the existence of a high velocity group, mainly
populated by late--type galaxies.

\subsection{Analysis of substructure}

We analyze the velocity distribution to look for possible deviations
from Gaussianity that could provide important signatures of complex
dynamics. For the following tests the null hypothesis is that the
velocity distribution is a single Gaussian.

We estimate three shape estimators, i.e. the kurtosis, the skewness,
and the scaled tail index (see, e.g., Beers et al. \cite{bee91}).  The
value of the normalized kurtosis (-0.71) and of the scaled tail index
(0.839) shows evidence that the velocity distribution differs from a
Gaussian, being lighter--tailed, with a c.l. of $\sim90-95\%$ (see
Table~2 of Bird \& Beers \cite{bir93}).

\begin{figure}
\centering 
\resizebox{\hsize}{!}{\includegraphics{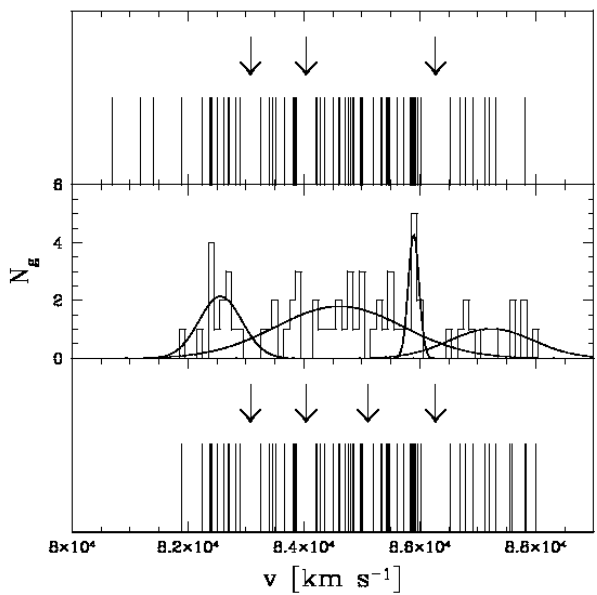}}
\caption
{Velocity distribution of radial velocities for the cluster
members. {\em Upper} and {\em lower panels}: 
stripe density plots where arrow indicate
the position of the significant gaps (samples--A and B, respectively). 
{\em Central panel}: velocity histogram with a binning of 100 \ks
with the Gaussians corresponding to the most significant KMM partition
for the sample--B.}
\label{figstrip}
\end{figure}

Then, we investigate the presence of gaps in the distribution 
using the ROSTAT package (Beers et al. \cite{bee90}).  A weighted gap
in the space of the ordered velocities is defined as the difference
between two contiguous velocities, weighted by the location of these
velocities -- precisely
the weight is $i\times({\rm N}_{\rm g}-i)$, where $i$ gives the
location of the object preceding the gap and ${\rm N}_{\rm g}$ is the
total number of galaxies. We obtain values for these gaps relative to
their average size, precisely the midmean of the weighted--gap
distribution. We look for normalized gaps larger than 2.25 since in
random draws of a Gaussian distribution they arise at most in about
$3\%$ of the cases, independent of the sample size (Wainer and Schacht
\cite{wai78}; see also Beers et al. \cite{bee91}). Three significant
gaps in the ordered velocity dataset are detected (see
Fig.~\ref{figstrip}).  In Table~\ref{tabgap} we list the number of
galaxies and the velocity of the object preceding the gap, the
normalized size (i.e., the ``importance'') of the gap itself, and the
probability of finding a normalized gap of this size and with the same
position in a normal distribution (as computed with ROSTAT package,
Beers et al. \cite{bee90}).  Using the results of weighted--gap
analysis, we divide the dataset in four subsets containing 17, 10, 32,
and 9 galaxies from low to high velocities.  We compare these subsets
two by two by applying the 2DKS--test to the galaxy positions. We find
no difference.

The velocity of the cD galaxy (main nucleus, ${\rm v}=84361$ \kss)
shows no evidence of peculiarity according to the Indicator test by
Gebhardt \& Beers (\cite{geb91}).

\begin{table}
        \caption[]{Results of weighted--gap analysis}
         \label{tabgap}
                $$
         \begin{array}{l c c c c}
            \hline
            \noalign{\smallskip}
            \hline
            \noalign{\smallskip}
\mathrm{Sample} & \mathrm{N_{gals,prec}} & \mathrm{v_{prec}}& \mathrm{Size}& \mathrm{Prob.}\\
                &                  &\mathrm{km\ s^{-1}}&              &
         \\
            \hline
            \noalign{\smallskip}
 
\mathrm{Sample\ A)} & 17 &82900&2.73& 1.4E-2\\
\mathrm{Sample\ A)} & 27 &83882&2.91& 6.0E-3\\
\mathrm{Sample\ A)} & 32 &86018&2.56& 1.4E-2\\
\mathrm{Sample\ B)} & 14 &82900&2.62& 1.4E-2\\
\mathrm{Sample\ B)} & 24 &83882&2.93& 6.0E-3\\
\mathrm{Sample\ B)} & 39 &85011&2.33& 3.0E-2\\
\mathrm{Sample\ B)} & 56 &86018&3.05& 2.0E-3\\
              \noalign{\smallskip}
            \hline
            \noalign{\smallskip}
            \hline
         \end{array}
$$
         \end{table}

The cluster velocity field may be influenced by the presence of
internal substructures.  To investigate the velocity field of A697, we
divide galaxies into low-- and high--velocity subsamples by using the
median value of galaxy velocities $\bar{\rm{v}}=84664$ \kss, and check
the difference between the spatial distributions of the two
samples. We find no difference between high-- and low--velocity
galaxies.  We also perform the multiple linear regression fit to the
observed velocities with respect to the galaxy positions in the plane
of the sky (see also Girardi et al.  \cite{gir96}; den Hartog \&
Katgert \cite{den96}), but we do not find any significant velocity
gradient.

We combine galaxy velocity and position information to compute the
$\Delta$--statistics devised by Dressler \& Schectman (\cite{dre88}).
This test is sensitive to spatially compact subsystems that have
either an average velocity that differs from the cluster mean, or a
velocity dispersion that differs from the global one, or both.  We
find $\Delta=69$ for the value of the parameter which gives the
cumulative deviation.  This value is not a significant indication of
substructure as assessed computing 1000 Monte Carlo simulations,
randomly shuffling the galaxy velocities.

Different substructure--tests are sensible to different kinds of
substructure (Pinkney et al. \cite{pin96}). Since above we find
evidence of substructure only from 1D tests we think that the possible
subclumps are (almost) LOS aligned.  Therefore, we attempt to detect
subsets in the velocity distribution resorting to the Kaye's mixture
model (KMM) test as implemented by Ashman et al. (\cite{ash94}).  The
KMM algorithm fits a user--specified number of Gaussian distributions
to a dataset and assesses the improvement of that fit over a single
Gaussian. In addition, it provides the maximum--likelihood estimate of
the unknown n--mode Gaussians and an assignment of objects into
groups. KMM is most appropriate in situations where theoretical and/or
empirical arguments indicate that a Gaussian model is reasonable.  The
Gaussian is valid in the case of cluster velocity distributions, where
gravitational interactions drive the system toward a relaxed
configuration with a Gaussian velocity distribution.  However, one of
the major uncertainties of this method is the optimal choice of the
number of groups for the partition.  We use the results of the gap
analysis to determine the first guess for the group partition and we
try to fit two, three or four velocity groups.  We do not find any
group partition which is a significant better descriptor of the
velocity distribution with respect to a single Gaussian.

\subsection{Further insights into substructure}

In this section we analyze the alternative sample of cluster members
(sample--B), which is likely more suitable to represent a
multi--Gaussians system.

The analysis of the alternative sample leads to a similar global
properties, i.e. $\left<{\rm v}\right>=(84790\pm$163) \kss,
$\sigma_{\rm v}=1349_{-63}^{+112}$ \kss, $M(<0.75
\hhh)=9.7_{-1.2}^{+1.8}$ \mquaa, and $M(<{\rm R}_{\rm vir}=3.89
\hhh)=4.6_{-1.2}^{+1.4}$ \mquii.  We also obtain similar results in
the substructure analysis in the sense that only velocity distribution
shows signs of possible substructure. In particular, the velocity
distribution shows a departure from Gaussian at the 98\%--99\%
c.l. according to the W--test (Shapiro \& Wilk \cite{sha65}) and the
value of kurtosis (-0.92). We also find an additional significant
weighted gap with respect to sample--A (see Fig.~\ref{figstrip}).
This additional gap is due to the fact that the slight change of
the sample (and of the consequent values of the weights, see
Sect.~3.4) slightly enhances the size of this gap from 2.21 to 2.33,
i.e. beyond our threshold of 2.25. In addition, the velocity of the
cD galaxy shows marginal evidence of peculiarity according to the
Indicator test by Gebhardt \& Beers (\cite{geb91}, at the 90--95$\%$
c.l.).

However, while in the sample--A KMM analysis fails in detecting a
significant partition, in the sample--B we find that each of the
2/3/4/5 group partitions is a better descriptor of the velocity
distribution at the $99.5\%$/$98.8\%$/$99.6\%$/$99.3\%$ c.ls.,
according to the likelihood ratio test.  The KMM results differ so
much between the sample--A and B that we investigated the possible
cause. The difference is likely due to the three lowest--velocity
galaxies in the sample--A: after rejecting these three galaxies, the
KMM analysis of sample--A give similar results to those of sample--B.

For each out of the 2/3/4/5 group partitions we use the KMM code to
assign members to the respective groups.  Table~\ref{tabkmm} lists the
results for the kinematical analysis. Note that the uncertainty in KMM
membership assignments leads to an artificial truncation of the tails
of the distributions and this may be leading to lower estimate for the
velocity dispersion (Bird \cite{bir94}).  Figure~\ref{figstrip} shows
the four group partition (hereafter KMM4g1, KMM4g2, KMM4g3, KMM4g4),
which is the most significant one.

\begin{table}
        \caption[]{Kinematical analysis for KMM groups}
         \label{tabkmm}
                $$
         \begin{array}{l r l l}
            \hline
            \noalign{\smallskip}
            \hline
            \noalign{\smallskip}
\mathrm{Sample} & \mathrm{N_g} & \phantom{249}\mathrm{<v>}\phantom{249} & 
\phantom{24}\sigma_{\rm v}^{\mathrm{a}}\phantom{24}\\
& &\phantom{249}\mathrm{km\ s^{-1}}\phantom{249} 
&\phantom{2}\mathrm{km\ s^{-1}}\phantom{24}\\
            \hline
            \noalign{\smallskip}
\hline
\mathrm{KMM2g1} & 12 &82545\pm\phantom{1}41 &\phantom{1}134_{-38}^{+20}\\
\mathrm{KMM2g2} & 57 &85273\pm148 &1110_{-91}^{+99}\\
\hline
\mathrm{KMM3g1} & 12 &82545\pm\phantom{1}41 &\phantom{1}134_{-38}^{+20}\\
\mathrm{KMM3g2} & 49 &84951\pm126 &\phantom{1}875_{-75}^{+101}\\
\mathrm{KMM3g3} &  8 &87557\pm104 &\phantom{1}260_{-30}^{+54}\\
\hline
\mathrm{KMM4g1} & 14 &82559\pm\phantom{1}56 &\phantom{1}198_{-42}^{+54}\\
\mathrm{KMM4g2} & 35 &84616\pm104 &\phantom{1}605_{-43}^{+64}\\
\mathrm{KMM4g3}^{\mathrm{b}} &  7 &85894\pm\phantom{1}19 &\phantom{10}51_{-82}^{+82}\\
\mathrm{KMM4g4} & 13 &87238\pm116 &\phantom{1}394_{-41}^{+69}\\
\hline
\mathrm{KMM5g1} & 14 &82559\pm\phantom{1}56 &\phantom{1}198_{-42}^{+54}\\
\mathrm{KMM5g2} & 10 &83682\pm\phantom{1}60 &\phantom{1}176_{-12}^{+68}\\
\mathrm{KMM5g3} & 24 &84949\pm\phantom{1}77 &\phantom{1}336_{-28}^{+42}\\
\mathrm{KMM5g4}^{\mathrm{b}} & 8  &85907\pm\phantom{1}19 &\phantom{10}51_{-82}^{+82}\\
\mathrm{KMM5g5} & 13 &87238\pm116 &\phantom{1}394_{-41}^{+69}\\
              \noalign{\smallskip}
            \hline
            \noalign{\smallskip}
            \hline
         \end{array}
$$
\begin{list}{}{}  
\item[$^{\mathrm{a}}$] We use the biweigth and the gapper estimators
by Beers et al. (1990) for samples with $\mathrm{N_g}\ge$ 15 and with
$\mathrm{N_g}<15$ galaxies, respectively (see also Girardi et
al. \cite{gir93}). 
\item[$^{\mathrm{b}}$] We list the value of
$\mathrm{\sigma_v}$ before the correction for the estimated errors in
the redshift measures 
since the correction leads to a negative value.
\end{list}
         \end{table}

Looking for correlations between positions and velocities, we compare
the KMM groups two by two by applying the 2DKS--test to the galaxy
position of member galaxies. In the four--groups partition we obtain
a marginal difference when comparing KMM4g3 with KMM4g1(KMM4g2) at the
$95.24\%$($91.12\%$) c.l..

KMM4g3 group (gals. with IDs.~4, 7, 16, 27, 31, 84, 93) has a very
small velocity dispersion, and shows a very sparse 2D distribution
with respect to the rest of the cluster.  Moreover, out of the six
KMM4g3 galaxies having $B$--$R$ colours, half of them belong to the
blue sample according to our definition in Sect.~3.3. Therefore,
this group might be connected with the high velocity group
suggested by our analysis in Sect.~3.3.

\section{2D galaxy distribution}

\begin{figure}
\centering
\includegraphics[width=7cm]{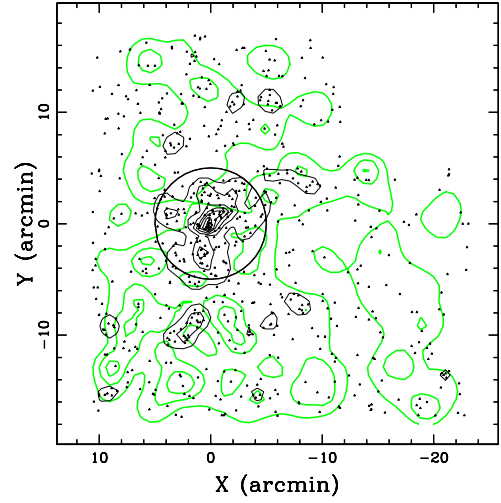}
\caption
{Spatial distribution on the sky and relative isodensity contour map
of 645 likely cluster members (according to the colour--magnitude
relation) with $R\le 21$, obtained with the adaptive--kernel method
(black lines).  For comparison is also shown the contour map of the 749
likely non-cluster members (grey lines).  The plot is
centered on the cluster center.  The circle indicate the 5\arcm
central region. }
\label{figk2A}
\end{figure}

By applying the 2D adaptive--kernel method to the position of A697
galaxy members we find only one peak.  However, our spectroscopic data
do not cover the entire cluster field and suffers for magnitude
incompleteness. To overcome these limits we recover our photometric
catalog.  We consider galaxies (objects with SExtractor stellar index
$\le 0.9$) lying within 0.25 mag from the CMR determined in Sect.~3.3.
The inspection of Fig.~\ref{figcm} suggests some contamination by
field galaxies already for $20\le R \le 21$ (cross symbols), thus we
do not show results for galaxies fainter than 21 mag.
Figure~\ref{figk2A} shows the contour map for 645 likely cluster
members having $R\le 21$: the main structure has a ``S--shape''
centered on the cD galaxy with a secondary, important NW peak. This
central structure is surrounded by few significant peaks within a
radius of 5\arcmm.  Another significant, dense peak in the galaxy
distribution lies $\sim9$\arcm south of the cluster center.

The field of A697 is strongly disturbed by the presence of several
luminous stars: this might bias our results about 2D cluster
morphology.  To face on this problem we consider the distribution of
galaxies having $R\le 21$ and lying outside 0.75 mag from the
colour--magnitude relation, i.e. likely mainly formed by non member
galaxies. Members and non members have rather different 2D
distributions suggesting that our results about cluster morphology are
not affected by the presence of disturbing luminous stars.

\begin{figure}
\centering
\includegraphics[width=7cm]{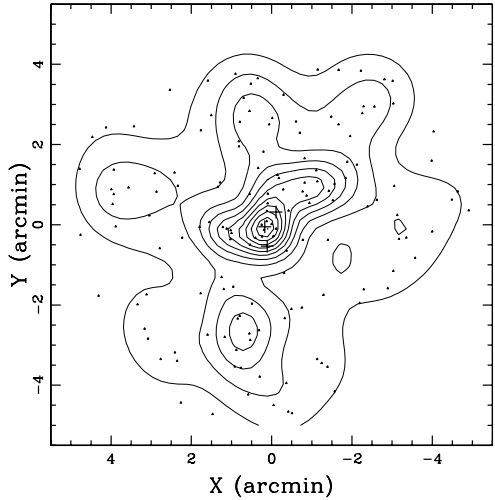}
\caption
{Spatial distribution on the sky and relative isodensity contour map
of 149 likely cluster members with $R\le 21$ within 1.3 \h
($\sim5$\arcmm) from the cluster center.  The plot is centered on the
cluster center.  Crosses indicate the position of the centers of the
substructures found by the wavelet technique in Chandra X--ray data
(see Sect.~5).}
\label{fig2113}
\end{figure}

\begin{figure}
\centering
\includegraphics[width=7cm]{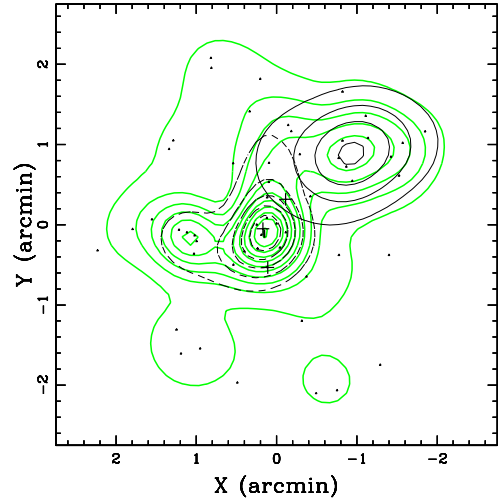}
\caption
{Spatial distribution on the sky and relative isodensity contour map
of 54 likely cluster members with $R\le 21$ within 0.6 \h from the
cluster center (gray lines). The contour maps of brilliant and faint
galaxies -- i.e. galaxies having $R \le20$ and $20< R\le 21$ -- are
also shown (black dashed and solid lines, respectively).  The plot is
centered on the cluster center.  Crosses indicate the position of the
centers of the substructures found through wavelet technique in
Chandra X--ray data (see Sect.~5).}
\label{figk2B}
\end{figure}

To definitely exclude the possibility of a biased cluster morphology
we also consider: a) the distribution of 149 likely members within a
radius of 1.3 \h ($\sim5$\arcmm) to avoid a very luminous star $\sim
7$\arcm East and a bad line originated by the gap between the CCD
chips of the WFC $\sim5$\arcm North (see Fig.~\ref{fig2113}); b) the
distribution of 54 likely members within a radius of 0.6 \h
($\sim2.5$\arcmm) to avoid a few luminous south--western stars (see
Fig.~\ref{figk2B}).  The first analysis finds again a southern peak:
the six galaxies of our spectroscopic catalog located within a radius
of 0.25 \h from its center
[R.A.=$08^{\mathrm{h}}43^{\mathrm{m}}01\dotsec1$, Dec.=$+36\degree
19\arcmm 18\arcs$ (J2000.0)] have a mean velocity comparable to that
of the whole system. Both a) and b) analyses confirm the S--shape of
the main central structure, but we have too  poor statistics to
analyze the kinematical properties of E and NW peaks.

Finally, we note that the NW peak is induced by the presence of faint
galaxies, while brilliant galaxies can be described by a single
central peak thus in agreement with the 2D analysis of the spectral
catalog.  Figure~\ref{figk2B} shows the contour maps of the
distributions of 32 brilliant ($R\le20$) and 22 faint ($20<R\le 21$)
galaxies within a radius of 0.6 \hh: the respective 2D distributions
are different at the 98.48\% c.l., according to the 2DKS test.
Suspecting that these faint galaxies belong to a background, dense
cluster we have studied the correspondent galaxies in our
spectroscopic sample, i.e.  the five non member galaxies having
$20<R\le 21$ and lying within 0.25 mag from the CMR (see crosses in
Fig.~\ref{figcm}). These five galaxies are scattered in redshift,
$z\in(0.29, 0.52)$, and in projected position around the cluster
center. Therefore, we have no evidence for a background NW cluster.

\section{X--ray data and analysis} 

\begin{figure}
\centering
\includegraphics[width=8cm]{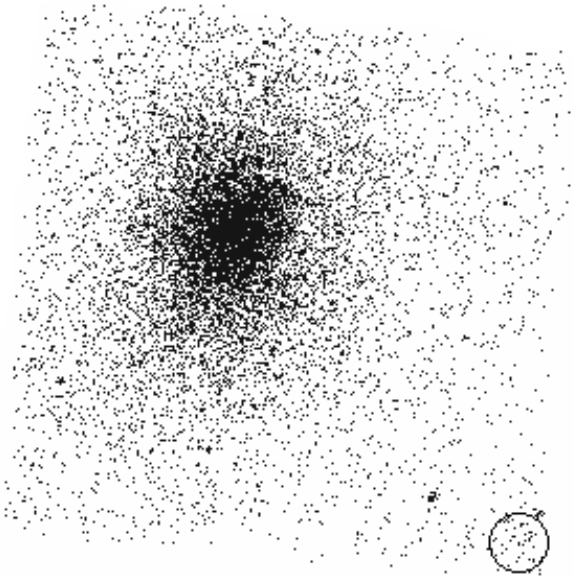}
\caption{8.5$\arcmm\,\times$\,8.5$\arcmm$ Chandra X--ray image (ID \#4217,
chip I3 only) of the cluster A697 in the energy band 0.5--2 keV. The
circle puts in evidence a faint diffuse emission located $\sim$6 \arcm
SW from the cluster center (see Sect.~5). North is at the top and East
to the left.}
\label{chipI3}
\end{figure}

The X--ray analysis of A697 is performed on the archival data of the
Chandra ACIS--I observation 800373 (exposure ID \#4217, see
Fig~\ref{chipI3}). The pointing has an exposure time of 19.8 ks. Data
reduction is performed using the package CIAO\footnote{CIAO is freely
available at http://asc.harvard.edu/ciao/} (Chandra Interactive
Analysis of Observations) on chip I3 (field of view
$\sim8.5\arcmm\times 8.5\arcmm$). First, we remove events from the
level 2 event list with a status not equal to zero and with grades
one, five and seven. Then, we select all events with energy between
0.3 and 10 keV. In addition, we clean bad offsets and examine the
data, filtering out bad columns and removing times when the count rate
exceeds three standard deviations from the mean count rate per 3.3 s
interval. We then clean I3 chip for flickering pixels, i.e., times
where a pixel has events in two sequential 3.3 s intervals. The
resulting exposure time for the reduced data is 19.5 ks.

In Fig.~\ref{figisofote} we plot an $R$--band image of the cluster
with superimposed the X--ray contour levels of the Chandra image. The shape
of the cluster appears to be moderately elliptical. By using the CIAO
package Sherpa we fit an elliptical 2D Beta model to the X--ray photon
distribution to quantify the departure from the spherical shape. The
model is defined as follows:
\begin{equation}
f(x,y)=f(\rm R)=A/[1+({\rm R}/{\rm R}_0)^2]^{\alpha}
\end{equation}
where the radial coordinate $\rm R$ is defined as ${\rm
R}(x,y)=[X^2(1-\epsilon)^2+Y^2]^{1/2}/(1-\epsilon)$,
$X=(x-x_0)\,\cos\,\theta+(y-y_0)\,\sin\,\theta$ and
$Y=(y-y_0)\,\cos\,\theta-(x-x_0)\,\sin\,\theta$. Here $x$ and $y$ are
physical pixel coordinates on chip I3. The best fit centroid position
is located on the main body of the cD galaxy. The best fit core
radius, the ellipticity and the position angle are ${\rm
R}_0$=49.3\arcs$\pm$3.5\arcs (i.e. 210$\pm$15 \kpcc), $\epsilon$=0.26$\pm$0.02
and PA=164.4$\pm$2.3 degrees (measured from North to East),
respectively.

To detect possible substructures in A697 we perform a wavelet analysis
by running the task CIAO/Wavdetect on chip I3  (for applications
of this technique in the literature see, e.g, Slezak et
al. \cite{sle94}; Vikhlinin et al. \cite{vik97}; Sun et
al. \cite{sun02}; Ferrari et al. \cite{ferra05}). The task was run on
different scales to search for substructures with different sizes.
The significance threshold\footnote{see \S~11.1 of the CIAO Detect
Manual (software release version 3.2, available at the WWW site
http://cxc.harvard.edu/ciao/manuals.html)} was set at $10^{-6}$.  The
results are shown in Fig.~\ref{figisofote}. Thick ellipses represent
three significant surface brightness peaks found by Wavdetect in the
core of the cluster. The principal one, located at
R.A.=$08^{\mathrm{h}}42^{\mathrm{m}}58\dotsec4$ and Dec.=$+36\degree
21\arcmm 56\arcs$, is centered 15$\arcs$ S--SE from the central cD
galaxy. Two more significant peaks are located at
R.A.=$08^{\mathrm{h}}42^{\mathrm{m}}58\dotsec1$ and Dec.=$+36\degree
21\arcmm 27\arcs$, and at
R.A.=$08^{\mathrm{h}}42^{\mathrm{m}}57\dotsec0$ and Dec.=$+36\degree
22\arcmm 18\arcs$, respectively. The wavelet technique reveals
the complex structure of the dense central cluster region, apparent in
the surface brightness distribution only when a very tuned choice of
the contour levels is used.

For the spectral analysis of the cluster X--ray photons, we compute a
global estimate of the ICM temperature. The temperature is computed
from the spectrum of the cluster within a circular aperture of 2\arcm
radius around the cluster center. Fixing the absorbing galactic
hydrogen column density at 3.41$\times$10$^{20}$ cm$^{-2}$, computed
from the HI maps by Dickey \& Lockman (\cite{dic90}), we fit a
Raymond--Smith (\cite{ray77}) spectrum using the CIAO package Sherpa
with a $\chi^{2}$ statistics. We find a best fitting temperature of
$T_{\rm X}=($10.2$\pm$0.8) keV and a metal abundance of 0.36$\pm$0.8
in solar units.

Very interestingly, the X--ray image (see Fig.~\ref{chipI3}) also
shows the presence of a faint diffuse emission located $\sim$6 \arcm
SW of the cluster center. Already revealed by M00 on a ROSAT HRI
image, this source could be an additional infalling
group. Unfortunately, we have no way to verify this hypothesis by
using redshift data and thus we do not discuss this feature in the following.

\section{Discussion}

We analyze the internal dynamics of A697 on the basis of spectroscopic
data for 93 galaxies in a cluster region of a radius of $\sim5$\arcm
(i.e., $\sim1.3$ \hh) from the cD galaxy.  We find that A697 appears
as a single peak in the redshift space at $\left<z\right>=0.282$.

Using a standard selection procedure we obtain 68 fiducial cluster
members (sample--A). We compute a LOS velocity dispersion of
$\sigma_{\rm V}=1334_{-95}^{+114}$ \kss, higher or comparable to the
highest values for other clusters in the literature (see Fadda et
al. \cite{fad96}; Mazure et al. \cite{maz96}; Girardi \& Mezzetti
\cite{gir01}). We obtain consistent estimates of $\sigma_{\rm V}$ --
see Table~\ref{tabv} -- when using alternative samples of cluster
members, i.e. the sample-B which is likely more suitable for a
multipeaked velocity distribution, the red--galaxies sample which is
likely less contaminated by field galaxies, and the sample obtained
after the rejection of galaxies belonging to the KMM4g3 group (possibly
infalling onto the cluster, see Sect.~6.2).  Moreover, our estimate of
$\sigma_{\rm v}$ is fully consistent with the average X--ray
temperature $T_{\rm X}=$10.2$\pm$0.8 keV coming from our analysis of
Chandra data when assuming the equipartition of energy density between
gas and galaxies (i.e. $\beta_{\rm spec} =1.06^{+0.20}_{-0.17}$ to be
compared with $\beta_{\rm spec}=1$\footnote{$\beta_{\rm
spec}=\sigma_{\rm v}^2/(kT/\mu m_{\rm p})$ with $\mu=0.58$ the mean
molecular weight and $m_{\rm p}$ the proton mass.}, see also
Fig.~\ref{figprof}).

Assuming that the cluster is in dynamical equilibrium and mass follows
galaxy distribution, we compute virial mass estimates obtaining
$M(<{\rm R}_{\rm max}=0.75 \hhh)=9.5^{+1.8}_{-1.5}$ \mqua and, using a
slightly different approach, $M(<{\rm R}_{\rm vir}=3.85
\hhh)=4.5_{-1.3}^{+1.4}$ \mqui for the region well sampled by data and
for the virialized region, respectively.  This makes A697 one of the
most massive clusters (e.g., Girardi \& Mezzetti \cite{gir01}), in
agreement with the results of Dahle et al. (\cite{dah02}).

To compare our result to the estimate obtained via gravitational
lensing we obtain a projected mass assuming that the cluster is
described by a King--like mass distribution (see Sect.~3.2) or,
alternatively, a NFW profile where the mass--dependent concentration
parameter is taken from Navarro et al. (\cite{nav97}) and rescaled by
the factor $1+z$ (Bullock et al. \cite{bul01}; Dolag et
al. \cite{dol04}).  We obtain $M_{\rm
proj}(<\rm{R}=150$\arcss)=(1.6--1.9)\mqui in agreement with that found
by gravitational lensing (Dahle et al. \cite{dah02}, after having
considered the different cosmological model).  Using the same mass
distributions we compute $M(<\rm{R}=1$ \hh)=(1.2--1.3)\mqui and, using
r--band luminosity by Popesso et al.~(\cite{pop04}), estimate
$M/L$=(300--330) \mll.  This value is comparable to those reported by
Carlberg et al.~(\cite{car96}) for clusters of similar redshift
contradicting some evidence for an unusual high mass--to--light ratio
(Dahle et al. \cite{dah02}).

\subsection{Internal structure}

The velocity dispersion profile rises in the central cluster region
out to $\sim0.5 $ \h (see Fig.~\ref{figprof}).  This explains why,
when analyzing 7(9) galaxies lying within $\sim 1$\arcm ($\sim 0.25$
\hh), M00 found somewhat smaller values of $\sigma_{\rm v}$ [=553(941)
\ks] with respect to our estimate.  Increasing profiles are sometimes
detected in clusters (e.g., den Hartog \& Katgert \cite{den96};
Girardi et al. \cite{gir98}) and might be the signature of a relaxed
cluster. In fact, the presence of circular velocities in a cluster
undergoing two--body relaxation in the central region leads to a rise
in the velocity dispersion profile (e.g., Merritt \cite{mer87};
Merritt \cite{mer88}; Sharples et al.  \cite{sha88}; Girardi et
al. \cite{gir98}). The same behaviour of the profile might be due to
the large efficiency of galaxy merging in the dense, central cluster
region (Menci \& Fusco Femiano \cite{men96}).  Alternatively,
increasing velocity dispersion profiles can be explained by the
presence of subclumps at different mean velocity (partially) LOS
aligned in the central cluster region.  As for A697, the latter
hypothesis is supported by the mean velocity profile which shows a
somewhat low value of $\left<{\rm v}\right>$ at about 0.5 \h (see
Fig.~\ref{figprof}, upper panel). Instead, as possible explanation, we
exclude the presence of still remaining field galaxies incorrectly
assigned to the cluster sample since the analysis of the red--galaxy
population, likely less contaminated by interlopers, leads to similar
$\sigma_{\rm v}$ and $\left<{\rm v}\right>$ profiles (see
Fig.~\ref{figprof}).

The presence of velocity and spatial segregation of galaxies with
respect to their colours, luminosities, and morphologies is often
taken as evidence of advanced dynamical evolution of the parent
cluster.  In fact, these segregation can be explained by the accretion
of subsequent generations of infalling field galaxies and/or secondary
relaxation phenomena such as dynamical friction (e.g., Biviano et
al. \cite{biv92}; Ellingson et al. \cite{ell01}; Biviano et
al. \cite{biv02}; Poggianti \cite{pog04}; Goto \cite{got05}).  We find
marginal evidence for the galaxy segregation as expected in a very
evolved cluster.  Rather, the difference we detect in the mean
velocity between the red and the blue populations suggests the
presence of a high--velocity group, mainly populated by late--type
galaxies. 

As concerning the presence of subgroups, we find evidence for the non
Gaussianity of the velocity distribution, but not for correlations
between galaxy positions and velocities, i.e. no significant velocity
gradients and Dressler--Schectman substructure. This suggests that the
possible subclumps are mainly LOS aligned (see also Pinkney et
al. \cite{pin96}). This LOS substructure is the likely cause of the
central rise of the velocity dispersion profile.

Unfortunately, our attempt to extract subgroups through the KMM
procedure does not lead to robust results.  In fact, although both the
velocity distributions of the samples-A and B have similar significant
weighted gaps, the KMM analysis gives significant results only in the
case of the sample--B. Moreover, for this sample we find few
alternative, significant partitions (i.e., 2/3/4/5 group partitions
being the 4-mode the most significant one).  In the case of A697, we
interpret the variety of these results in the sense that the presence
of LOS substructure is real, but we look with caution at the
quantitative results of the KMM analysis without an external support
coming from the correlation between positions and velocities.  We find
a marginal correlation only in the case of the KMM4g3 group (at
$\left<{\rm v}\right>=85894$ \kss, see Sec.~3.4). This group has a
very small velocity dispersion, $\sigma_{\rm v}<100$ \kss, and a
sparse galaxy distribution thus resembling characteristics of a loose
group (e.g., Geller \& Huchra \cite{gel83}; Ramella et
al. \cite{ram89}; Giuricin et al. \cite{giu00}) likely still forming
(e.g., Giuricin et al. \cite{giu88}; Diaferio et al. \cite{dia93}),
rather than of a relaxed core of a secondary cluster, in agreement
with its large fraction of blue galaxy population.  More in general,
all the 2/3/4/5 group partitions detect the presence of a low velocity
group, $\left<{\rm v}\right>\sim85500$ \kss, with a velocity
dispersion of $\sigma_{\rm v}=$100--200 \ks and all the 3/4/5 group
partitions detect the presence of a high velocity group, $\left<{\rm
v}\right>\sim87500$ \kss, with $\sigma_{\rm v}=$200-400 \kss.  The
size of the main group depends on the mode of the partition
($\sigma_{\rm v}=$300--1100 \kss), being about the typical
cluster--mass in the most significant four--group partition
($\sigma_{\rm v}\sim 600$ \kss).

Analysis of Chandra data give us other indications that A697 is far
from being fully relaxed.  According to the X-ray emission, the shape
of the cluster appears to be elliptical ($\epsilon=$0.26$\pm$0.02),
elongated in the SSE--NNW direction (PA=164.4$\pm$2.3 degrees).  The
value of ellipticity is moderate, comparable to the median values
recovered in wide cluster samples (e.g., Mohr et al. \cite{moh95}; De
Filippis et al. \cite{def05}).  The direction of the elongation agrees
with previous results from ROSAT and Chandra (M00; De Filippis
et al. \cite{def05}; see also ROSAT HRI image by Ota \& Mitsuda
\cite{ota04}) and agrees with that of the inner cD isophotes (PA$=163$
degrees; M00) and the mass distribution recovered from gravitational
lensing (see upper right plot of Fig.~24 of Dahle et
al. \cite{dah02}). We find that the isodensity contour map
of the galaxy distribution shows an elongation toward NW in the
central cluster region, too (see Sect.~4). Interestingly, radio contour map
also shows a NW elongation in both WENSS and NVSS radio images
(Kempner \& Sarazin \cite{kem01}, see also Fig.~\ref{figisofote}).

The new, interesting result of our X--ray study is the presence of
three structures detected via wavelet analysis.  The main one almost
coincides with the cD galaxy and thus corresponds to the cluster core.
The other two -- one at NW and the other at south -- are likely to be
related with subgroups located in (or projected onto) the very central
cluster region, roughly within 0.2 \h from the cluster center.  The NW
X--ray substructure coincides with an unusual, low surface brightness
extended feature, the colour of which is consistent with that of an
old stellar population at this redshift (M00, see also our
Fig.~\ref{figimage}).  The 2D analysis of galaxy distribution using
the photometric catalog also indicates the presence of small structures
in the central cluster region, but in a somewhat larger scale, roughly
within 0.4 \h from the cluster center. Finally, we note that X--ray 
structures are not perfectly aligned along the direction
of the main axis of elliptical X--ray emission, rather they partially recover 
(on a smaller scale) the S--shape of the 2D galaxy distribution.

\subsection{Merging scenario}

On the basis of the asymmetric cD profile, extended features, and
highly elliptical potential implied by the arc model, M00 suggested
that or A697 has recently undergone a significant merger event or that
its cD is undergoing the process of forming its extended halo.

The present study supercedes the results of M00 finding evidence for
substructure in the velocity distribution, in the X--ray emission, and
2D galaxy distribution. Moreover, although we confirm previous
literature about the moderate value of ellipticity in X--ray emission,
so many cluster components show an elongation toward the NW direction
that the cluster asphericity is likely to be connected to an
important, large--scale physical phenomenon such as a cluster merger.
In particular, the evidence for subclumps in the velocity
distribution, the cluster elongation, and the variety of substructures
we detect lead us to conclude that A697 suffers of a multiple merger
event occurring roughly mainly along the LOS, with a transverse
component in the SSE--NNW direction.

The importance of the merging is still an open question.  In fact, we
do not detect large size substructures, we do not find any
large--scale correlation between positions and velocities of member
galaxies, and we are not able to unequivocally detect subclumps via
KMM procedure.  Therefore, we might be looking at very small clumps
(e.g., 0.2 \h the typical size of small--scale substructures, Girardi
et al. \cite{gir97}) accreted by a very massive cluster or,
alternatively, at the remnants of a very old merger (e.g., the
elongation of the X--ray emission is a long--lived phenomenon, up to 5
Gyr after the core passage; see Roettiger et al. \cite{roe96}).  In
both the cases, one would expect a good agreement between X--ray
temperature and velocity dispersion as we observe.

Alternatively, the merging might be more important/recent, but the
analysis is complicated because the merging axis is very close to the
LOS.  This would explain the absence of correlation between galaxy
positions and velocities, while the small--size X--ray substructures
might be identified with the cores of merging systems.  The absence of
a cooling core supports the hypothesis that the cluster was undergone
to important merging phenomena (Bauer et al. \cite{bau05}).  

Considering the face on most significant result of the KMM analysis
(KMM4g), the merger would concern -- or would have concerned --
a typical size cluster, $\sigma_{\rm v}\sim600$ \kss, with two smaller
systems, $\sigma_{\rm v}\sim$200--400 \kss, maybe detected as the two
X--ray substructures.  The low--$\sigma_{\rm v}$, loose group KMM4g3
would be still infalling onto the cluster since only very dense galaxy
structures are destined to survive to a merger event
(Gonz\'alez--Casado et al. \cite{gon94}).  However, a multiple
merger of systems with small, roughly comparable $\sigma_{\rm v}$
would lead to a much less massive cluster since the cluster mass
contained within R$_{\rm vir}$ goes as $\sigma^3_{\rm v}$ and that
contained within a fixed radius in Mpc goes as $\sigma^2_{\rm
v}$. This would be incompatible with the reasonable value of $M/L$ and
$\beta_{\rm spec}$ we find (but see Schindler \& M\"uller
\cite{sch93} for the enhancement of $T_{\rm X}$ during a cluster
merger). Moreover, in the case of A2744, the KMM analysis is far from giving
robust results. In fact, the two--group partition, which is also very
significant, suggests a quite different scenario where a very massive
cluster, $\sigma_{\rm v}\sim 1100$ \kss, accretes a much smaller
system, $\sigma_{\rm v}\sim 100$ \kss.

To obtain further insights into the importance and phase of the
merging event, as well as to compute a more reliable mass estimate one
would need to have more data and to sample the cluster at much larger
radii.  In fact, our sampled field is not satisfactory for such
unusual, very massive cluster ($\rm{R}_{\rm out}=0.34\times
\rm{R}_{\rm vir}$).  Indeed, the situation of A697 might be comparable
to that of the nearby cluster A2670 ($z\sim0.077$) which likes A697 as
concerning many features, e.g. the rise of the velocity dispersion
profile in the central region (Sharples et al.  \cite{sha88}) and the
absence of obvious X--ray substructure (as shown by an {\it Einstein}
image, but see recent Chandra results by Fujita et al. \cite{fuj06}).
Although most of A2670 properties suggested that it is a rich, relaxed
cluster, following analyses showed that it may consists of subclusters
that are merging along the LOS. However, this evidence only rose
sampling the full virial region of the cluster with two hundreds of
galaxies (Bird \cite{bir94}; Hobbs \& Willmore \cite{hob97}).

Finally, it is possible that the cD is forming with the accretion of
new mass when new groups collide with the cluster (as shown by
cosmological numerical simulations, G. Murante private comm.).  For
instance, the NW optical extended feature corresponds to a well
defined X--ray structure along the direction of the present cluster
elongation suggesting a connection with the present merger.  The
(likely) secondary nucleus in the NE direction might be connected with
an older merger.  This old merger might have formed the present main
cluster and, in fact, in all our KMM results both cD nuclei are
located in the same structure.

\section{Summary \& conclusions}

We present the results of the dynamical analysis of the rich, X--ray
luminous, and hot cluster of galaxies A697, likely containing a
diffuse radio emission.

Our analysis is based on new redshift data for 93 galaxies, measured
from spectra obtained at the TNG, in a cluster region within a radius
of $\sim1.3$ \h from the cluster center. We also use new photometric
data obtained at the INT telescope in a 30\arcmm$\times$30\arcm field.

We find that A697 appears as a peak in the redshift space at
$\left<z\right>=0.282$, which includes 68 galaxies recognized as
cluster members.  We compute the line--of--sight (LOS) velocity
dispersion of galaxies, $\sigma_{\rm v}=1334_{-95}^{+114}$ \kss, in
agreement with the high average X--ray temperature $T_{\rm
X}=($10.2$\pm$0.8) keV recovered from our analysis of Chandra data, as
expected in the case of energy--density equipartition between galaxies
and gas.  Assuming cluster is in dynamical equilibrium and mass
follows galaxy distribution, the virial theorem leads to $M(<{\rm
R}_{\rm max}=0.75 \hhh)=9.5^{+1.8}_{-1.5}$ \mqua and $M(<{\rm R}_{\rm
vir}=3.85 \hhh)=4.5_{-1.3}^{+1.4}$ \mqui for the region well sampled
by spectroscopic data and for the virial region, respectively.

Further insights show evidence of non complete dynamical relaxation:

\begin{itemize}

\item the non Gaussianity of the velocity distribution according to
different tests;

\item 
the presence of small--size substructures in the central cluster
region as shown by the X--ray emission and 2D galaxy distribution;

\item the elongation of X--ray emission
(PA=164.4$\pm$2.3 degrees) in agreement with the above
2D distribution and other cluster components as shown in the literature
(total mass, cD envelope, radio image);

\end{itemize}

We conclude that A697 suffers for a complex merger event occurring
roughly mainly along the LOS, with a transverse component in the
SSE--NNW direction.

That the KMM procedure gives no robust results, as well as that we
detect only small size substructures and do not find any large--scale
correlation between positions and velocities of member galaxies
suggest that we might be looking at very small clumps accreted by a
very massive cluster or at the remnants of a very old merger.
Possibly, the merging process might be more important/recent, but
the analysis is complicated because the merging axis is very close to
the LOS.  To obtain further insights into the importance and phase of
the merging event one would need to have more data and sample the
cluster at much larger radii.

The spatial correlation between the (likely) radio halo and the X--ray
and optical cluster structure supports the hypothesis of the
connection between radio extended radio emission and merging
phenomena.

\begin{acknowledgements}
We thank Joshua A. Kempner for helpful suggestions.  We thank Andrea
Biviano, Giuseppe Murante, and Mario Nonino for interesting
discussions. We thank the referee, Hans B\"ohringer, for his useful 
suggestions.\\
This publication is based on observations made on the island of La
Palma with the Italian Telescopio Nazionale Galileo (TNG), operated by
the Fundaci\'on Galileo Galilei - INAF (Istituto Nazionale di
Astrofisica), and with the Isaac Newton Telescope (INT), operated by
the Isaac Newton Group (ING), in the Spanish Observatorio of the Roque
de Los Muchachos of the Instituto de Astrofisica de Canarias.\\
This publication also makes use of data obtained from the Chandra data
archive at the NASA Chandra X--ray center (http://asc.harvard.edu/cda/).
\end{acknowledgements}

\end{document}